\documentclass[a4paper,10pt]{article}
\pdfoutput=1

\usepackage{devanagari}

\def\as(#1){{\alpha_{\rm s}^{\,#1}}}

\usepackage{
graphicx,
amsmath,
amssymb,
charter,
xcolor,
ifluatex,
booktabs,
bbold}
\usepackage{subfigure}     
\usepackage{colortbl}

\definecolor{Gray}{gray}{0.95}
\definecolor{RGray}{gray}{0.85}
\definecolor{CGray}{gray}{0.92}

\definecolor{tit}{rgb}{0.1,0.2,0.4}
\definecolor{blus}{cmyk}{1,1,0,0.6}
\definecolor{verde}{cmyk}{0.92,0,0.59,0.25}
        
 \allowdisplaybreaks          

\usepackage{tipa}
\usepackage{array}
\usepackage{amsmath,amssymb,amsfonts, bm}
\usepackage{dsfont}
\usepackage{slashed}
\usepackage{color}

\usepackage{caption}
\usepackage{commath}
\usepackage{calc}

\usepackage{latexsym}
\usepackage{slashed}
\usepackage{hyperref}
\usepackage{tikz}
\usepackage{axodraw2}

\usepackage{color,colordvi}

\newcommand{\be}{\begin{equation}}
\newcommand{\ee}{\end{equation}}

\newcommand{\bea}{\begin{eqnarray}}
\newcommand{\eea}{\end{eqnarray}}

\newcommand{\bfig}{\begin{figure}}
\newcommand{\efig}{\end{figure}}

\newcommand{\lmM}{l_{\mu M}}

\usepackage{commath}
\usepackage{calc}

\newcommand{\beq}{\begin{equation}}
\newcommand{\eeq}{\end{equation}}

\newcommand{\MSb}{$\overline{\mbox{MS}}$}

\newcommand{\als}{\alpha_{\rm s}}
\newcommand{\ars}{a_{\rm s}}

\usepackage{array}

\makeatletter
\newcommand*{\rom}[1]{\expandafter\@slowromancap\romannumeral #1@}
\makeatother

\usepackage{float}

\begin{document}

\allowdisplaybreaks
\vspace*{-2.5cm}
\begin{flushright}
{\small
IIT-BHU
}
\end{flushright}

\vspace{2cm}

\begin{center}
{\LARGE \bf \color{tit}  Renormalization-group improved Higgs to two gluons decay rate }\\[1cm]

{\large\bf Gauhar Abbas$^{a}$\footnote{email: gauhar.phy@iitbhu.ac.in} \\ 
%Saurabh Rindani$^{e}$\footnote{   } \\
Astha Jain$^{a}$\footnote{email: astha.jain.phy16@itbhu.ac.in    } \\
Vartika   Singh$^{a}$\footnote{email: vartikasingh.rs.phy19@itbhu.ac.in}   \\
Neelam  Singh$^{a}$\footnote{email: neelamsingh.rs.phy19@itbhu.ac.in  }  }  
\\[7mm]
{\it $^a$ } {\em Department of Physics, Indian Institute of Technology (BHU), Varanasi 221005, India}\\[3mm]
%{\it $^b$  } {\em }\\[3mm]

\vspace{1cm}
{\large\bf\color{blus} Abstract}
\begin{quote}
We  investigate the renormalization-group scale and scheme dependence of the  $H \rightarrow gg$   decay rate at the order N$^4$LO in the renormalization-group summed perturbative theory,  which employs the  summation of all renormalization-group accessible logarithms including the leading and subsequent four sub-leading logarithmic contributions to the full perturbative series expansion.     Moreover,  we study the higher-order behaviour  of the  $H \rightarrow gg$   decay width using the asymptotic Pad\'e approximant method in four different renormalization schemes.  Furthermore,  the higher-order behaviour is independently investigated in the framework of the asymptotic Pad\'e-Borel  approximant method where generalized Borel-transform is used as an analytic continuation of the original perturbative expansion.  The predictions of the asymptotic Pad\'e-Borel  approximant method are  found to be in agreement with that of the asymptotic Pad\'e approximant method. Finally,  we provide the $H \rightarrow gg$   decay rate at the order N$^5$LO  in the fixed-order  $ \Gamma_{\rm N^5LO} \,=\, \Gamma_0 (1.8375 \pm  0.047 _{\alpha_s(M_Z),1\%}\pm 0.0004_{M_t} \pm 0.0066_{M_H} 
  \pm 0.0036_{\rm P} \pm 0.007_{\text{s}}   \pm 0.0005_{sc} ),$ and  $\Gamma_{\rm RGSN^5LO} \,=\,  \Gamma_0 (1.841 \pm  0.047 _{\alpha_s(M_Z),1\%} \pm  0.0005_{M_t}\pm 0.0066_{M_H} \pm 0.0002_{\mu} \pm 0.0027_{\rm P}  \pm 0.001_{sc} )$  in the renormalization-group summed perturbative theories. 
\end{quote}

\thispagestyle{empty}
\end{center}

\begin{quote}
{\large\noindent\color{blus} 
}

\end{quote}

\newpage
\setcounter{footnote}{0}
%\maketitle
%\flushbottom

\def\ca{{C^{}_{\!A}}}
\def\cf{{C^{}_F}}

\section{Introduction}
The discovery of the Higgs boson at the Large Hadron Collider  (LHC) is the first step towards a better understanding of the electroweak symmetry breaking of the standard model (SM) \cite{atlas,cms}.   Moreover, studying the phenomenological behaviour of this particle is immensely important for providing  crucial information for  physics beyond the SM.  In the SM, the Higgs boson dominantly decays to a pair of bottom quarks,  that is,  $H \rightarrow \bar{b} b \rm (+ hadrons)$  followed by the  second leading hadronic decay channel  $H \rightarrow gg$, which   is predominantly mediated by the top quark.

There has been an extensive theoretical investigation of the  decay  $H \rightarrow gg$ in literature\cite{Ellis:1975ap,gghlo,Spira:1995rr}.  For instance,  it has been computed in an effective theory approach by considering $M_H \ll 2 m_t$ in reference \cite{Inami:1982xt}.  The  1/$m_{t}$  corrections are evaluated up to N$^2$LO in references \cite{Larin:1995sq,Schreck:2007um}.   The effective Higgs coupling to gluons is computed up to N$^4$LO in references \cite{KLS96,dec-as3,dec-as4a,dec-as4b,Kniehl:2006bg,ChBK-LL16,Gerlach:2018hen},   up to N$^3$LO in the limit of heavy top quarks in references \cite{gghnlolim, gghnnlo,gghn3lo},  and up to NLO with  the full quark mass dependence in references \cite{gghnlo,gghnlo0}.   The subleading  finite top-quark mass effects  in the large top mass expansion are known at N$^2$LO \cite{gghnnlomt},  and  the top-bottom interference  at NLO accuracy is computed in reference \cite{Mueller:2015lrx,Lindert:2017pky}.

We observe that  a low energy theorem can also yield the QCD corrections to the $H \rightarrow gg$ decay width at N$^4$LO  in the heavy quark limit \cite{Ellis:1975ap,Shifman:1979eb}.  The effective  Higgs-gluons Lagrangian in the heavy top-quark limit in this approach is derived by keeping the terms that depend only on the Higgs field, and using the transformation of the gluonic field strength operator and the strong coupling constant from $n_f +1$ to $n_f$ flavours \cite{Spira:2016zna,Spira:2016ztx}.

The absorptive part of the corresponding vacuum polarization is computed up to  N$^2$LO in references \cite{Inami:1982xt,Djouadi:1991tka,HggNNLO}.  The N$^3$LO corrections to the absorptive part  have been obtained in references \cite{HggN3LO,MV2,DaviesSW17,Davies:2017rle}.  Moreover,  the $H \rightarrow gg$ decay rate  is computed in the limit of a heavy top quark and any number of massless light flavours at N$^4$LO in reference~\cite{Herzog:2017dtz}.  The new theoretical development in this work is the computation of the absorptive part of the corresponding vacuum polarization at the order  N$^4$LO,  which was  missing so far.   The  computed correction to  the $H \rightarrow gg$ decay rate at N$^4$LO is slightly smaller than the $1/m_{top}$ effect at N$^2$LO,  and  larger than the presently unknown $1/m_{top}$ effect at N$^3$LO~\cite{Herzog:2017dtz}.  Moreover,  this computation has reduced the  uncertainty  due to the truncation of the perturbation series,  thus providing an improvement  in the $H \rightarrow gg$ decay rate.

The hadronic processes such as  $H \rightarrow gg$ are usually calculated up to finite order within perturbative QCD.  Therefore, these calculations are plagued by a significant dependence on the  renormalization (RG) scale parameter  $\mu$.   There are several prescriptions for dealing with this dependence in literature so that a physical prediction can be obtained.  For instance,  this scale in practice,  can be identified with the mass of the decaying particle.  One possibility is   to use a value of $\mu$ such that the known calculated terms in perturbation expansion show a local insensitivity to this value.  In addition,  a value of  $\mu$ may also be chosen,  which minimizes the highest-order known term of the perturbative expansion \cite{Chishtie:2000hg}.   It is obvious that using different prescriptions may cause  theoretical uncertainty in a prediction.  The ambiguity arising due to the RG scale dependence in $H \rightarrow gg$ decay at the order  N$^4$LO is of particular interest  for the precision physics within the SM as well as beyond the SM physics~ \cite{Chishtie:2000hg,Gao:2021wjn,Zeng:2018jzf,Elias:2000iw}.

We note that the renormalization group equation (RGE) is extensively used as a tool to improve  the behaviour of the perturbative expansions,  and extend their domain of applicability in perturbative quantum field theory.   For instance,  if the perturbative expansion is known to $k$-subleading orders,  the renormalization group equation can be used to sum the leading and subsequent  $(k-1)$ subleading  log contributions  to the full perturbative expansion.  Such logarithms are known as ``RG-accessible" in literature\cite{Ahmady1,Ahmady2}.

In this work,  we discuss the closed-form summation of  these RG-accessible logarithms  for  the   $H \rightarrow gg$  decay rate in the so-called ``renormalization-group summed perturbation theory" (RGSPT) in the four different RG schemes,  namely,   \MSb\,, SI, OS, and miniMOM. Perturbative series in the  RGSPT is  expanded in terms of coupling constant and logarithms such that  coefficient of every such term can be determined through RG-invariance in terms of its leading coefficient.  This is a generalization of the method of the leading logarithm summation  and provides the RG-summed perturbative expansion of the series to any given order of the perturbation theory.    One of the salient features of the new RGSPT expansions is the reduced sensitivity to  RG scale $\mu$ in spite of the presence of large logarithms \cite{Ahmady1}.

In literature,  RGE is found to be helpful  in the extraction of the divergent parts of the bare parameters using the determination of higher-order terms \cite{tHooft:1973mfk,Collins:1973yy}.  However, the  incorporation of all available higher-order RG-accessible terms to a given order in perturbation theory was first discussed by Maxwell  to deal with unphysical RG scale dependence \cite{Maxwell}.   This was further applied to moments of QCD lepto-production structure functions, and to  N$^2$LO correlation functions for the summation of the leading logarithms in references \cite{MaMi1,MaMi2}.    This  formalism  is also used by McKeon to extract  one-loop RG functions of $\phi^4$- and $\phi^6$-field theories  by performing the summation of leading logarithms to all orders \cite{McKeon:1998tr}.  The summation procedure of Mckeon was extended  to study the semileptonic $B$-decay rate  in reference \cite{Ahmady1}.

The RGSPT expansions have been  employed in various decays and observables in literature, and are shown to exhibit a remarkable improvement in the sensitivity  to the renormalization scale.  For instance,  these are used to study  the  $e^+e^-$ hadronic cross-section \cite{Ahmady2},  extraction of the strong coupling constant from the hadronic width of the $\tau$-decays \cite{Abbas:2012fi,Abbas:2012py},   extraction of the strange quark mass from moments of the $\tau$-decay spectral function \cite{Ananthanarayan:2016kll} and investigate the QCD static energy at the four-loop in reference~\cite{Ananthanarayan:2020umo}.  Moreover,  RGSPT expansions can  further be improved by the method of conformal mapping \cite{Abbas:2012fi,Abbas:2012py,Abbas:2016qmf,Abbas:2013usa,Abbas:2013eba,Abbas:2013awr,Caprini:2013bba,Caprini:2017ikn}.

In addition to the RG improvement of the  $H \rightarrow gg$  decay rate,   we also investigate  the  higher-order behaviour  of the perturbative expansion of  the  $H \rightarrow gg$  decay rate using the asymptotic Pad\'e approximant (APAP) method~\cite{Samuel:1992xd,Samuel:1994jv,Samuel:1995jc,Ellis:1996zn,Ellis:1995jv,Ellis:1997sb,Chetyrkin:1997dh,vanRitbergen:1997va,Elias:1998bi}.   Moreover,  the higher-order estimate of the beta and gamma function coefficients are also discussed using the same method.    These predictions are further independently obtained through the asymptotic Pad\'e-Borel  approximant (PBA) method, which provides an additional test of the  asymptotic Pad\'e approximant predictions.

Our paper will be presented along the following track: We first briefly review the theoretical formalism,  and state-of-the-art computation of the  $H \rightarrow gg$  decay rate at  N$^4$LO  in section \ref{sec2}.   This is followed by section \ref{sec3} where we discuss the $H \rightarrow gg$  decay width in the standard ``fixed-order perturbation theory" (FOPT),  and extract the expansion coefficients in various RG schemes.  In section \ref{sec4},  we discuss the $H \rightarrow gg$  decay rate  in the RGSPT at  N$^4$LO.    The expansion functions of the RGSPT are derived in an analytic closed-form by solving  the differential equation using  the method of iteration in this section.  We present a thorough discussion of the scale and scheme dependence of the  $H \rightarrow gg$  decay rate at  N$^4$LO  in the FOPT  in section \ref{SSDF}.  This is followed by the discussion of scale and scheme dependence of  the $H \rightarrow gg$  decay rate at  N$^4$LO in the RGSPT in section \ref{SSDR}.  The higher-order behaviour  of the  $H \rightarrow gg$  decay rate using the APAP formalism is discussed in section \ref{APAP}.  The study of the higher-order corrections to the  $H \rightarrow gg$  decay rate in the framework of the PBA is presented in section \ref{B-APAP}.   We determine  the $H \rightarrow gg$ decay width using the FOPT and the new RGSPT expansions discussed in this work in section \ref{sec6}  in the \MSb\,,   scale invariant (SI),   on shell    (OS),   and the minimal momentum subtraction (miniMOM) schemes.    We summarize our results in section \ref{sec7}.
\section{Inclusive decay of the  Higgs-boson to gluons}
\label{sec2}
In this section,  we briefly review the calculation of the  inclusive decay of the  Higgs-boson to two gluons.    The inclusive decay of the  Higgs-boson to gluons is calculated in the limit of a large top-quark mass with effectively massless $n_f$ flavours using  the following effective Lagrangian \cite{Inami:1982xt,HggNNLO},
\beq
\label{eq:Leff}
  \mathcal{L}_{\mathrm{eff}}  = \mathcal{L}_{\mathrm{QCD}(n_f)} \,-\, 2^{1/4\,}  G_{F}^{1/2} C_1 H G^{\mu\nu}_a G_{\mu\nu}^{a},
\eeq
where $H$  represents the Higgs field,  the renormalized gluon field-strength tensor  is denoted by  $G^{\mu\nu}_a$  with  $n_f$ flavours, the Lagrangian $ \mathcal{L}_{\mathrm{QCD}(n_f)} $ is the QCD Lagrangian, the  renormalized  coefficient $C_1$ parametrizes  the top-mass dependence,  and $G_{\rm F}$ is  the Fermi  coupling constant. 

The partial decay width of  the  Higgs-boson to gluons $\Gamma_{H \to\,gg}$ decay  can be computed using  the imaginary part of the Higgs-boson self-energy and  written as,
\beq
\label{GamHgg}
  \Gamma_{H \rightarrow  gg} =  \frac{\sqrt{2}\, G_{\rm F}}{M_H}\: |C_1|^2 \,
  \mathrm{Im}\, \Pi^{GG}(-M_H^2-i\delta),
\eeq
where $M_H$ is the mass of the  Higgs boson, $\Pi^{GG}$ represents the contribution to the self-energy of the Higgs boson due to its effective coupling to gluons,   and  $\delta$ is a positive real parameter,  which is infinitesimally small.

The  coefficient  $C_1$ is known up to N$^4$LO  and  its perturbative expansion is given by,
\begin{align}
\label{C1exp} 
  C_{1} & = - \frac{1}{3}\, \ars  \left(  1 + \sum_{n=1} \, c_{n} \, a_s^n (\mu^2) \right),
\end{align}
where $ \ars  \equiv  \frac{\als^{n_f}}{4 \pi}$ where $n_f$ are number of light flavours.

The known expansion coefficients $c_n$ in the SI scheme at the renormalization scale $\mu= \mu_t$,  where $\mu_t = m_t (\mu_t)$ is the   \MSb\  top quark mass evaluated at scale $\mu_t$,  in the \MSb\,   and  OS schemes  are written as\cite{Herzog:2017dtz},
\begin{align}
\label{C1gg}
 c_1  = &\,11, \
  c_2 =\,154.278 + 19 L_t +\left(-11.1667 + 5.33333 L_t\right)n_f,  \\ \nonumber 
  c_{3,\,\text{SI}}=&\,3031.7 + 537.111 L_t + 
 209 L_t^2 + (-492.139 + 107.852 L_t + 46 L_t^2) n_f\\ \nonumber & + (-14.1255 +  2.85185 L_t - 3.55556 L_t^2) n_f^2\\ \nonumber
 c_{4,\,\text{SI}}=&\,79815.7 + 12340.2 L_t + 9831.33 L_t^2 + 2299 L_t^3\\ \nonumber &+(-15966.1 - 1482.59 L_t + 1394.11 L_t^2 + 366.667 L_t^3) n_f\\ \nonumber &+(413.572 - 621.891 L_t - 94.5741 L_t^2 - 69.7778 L_t^3) n_f^2\\ \nonumber &+(-8.79068 + 23.7531 L_t - 2.85185 L_t^2 + 2.37037 L_t^3) n_f^3\\ \nonumber
c_{3,\,\overline{\text{MS}}} =&\,c_{3,\text{SI}} - 152L_t - 42.6667 L_t n_f\\ \nonumber
c_{4,\,\overline{\text{MS}}} =& \, c_{4,\text{SI}}  - 5576.22 L_t - 
 4230.67 L_t^2 + (-1158.59 L_t - 934.222 L_t^2) n_f + (-5.03704 L_t + 
    71.1111 L_t^2) n_f^2 \\ \nonumber
c_{3,\,\text{OS}}=&\,c_{3,\,\text{SI}}+202.667 + 56.8889 n_f\\ \nonumber
c_{4,\,\text{OS}}=&\, c_{4,\,\text{SI}}+13362.3 + 6688L_t + (2659.9 + 1472 L_t) n_f + (-147.307 - 113.778 L_t) n_f^2,
\end{align}
where $L_t= \ln (\mu^2/m_t^2)$, $\mu$ is the RG scale and $m_t$ is the definition of the top quark mass in the corresponding RG scheme.

The absorptive part of the vacuum polarization is  computed at  N$^4$LO  in reference \cite{Herzog:2017dtz},  thus,  making  the Higgs-boson to gluons decay width complete at the  N$^4$LO order.  The absorptive part of the vacuum polarization is written in the following form,
\bea
\label{ImGGexp}
  \frac{4\pi}{N_{ A} q^4}\, \mbox{Im}\, \Pi^{\,GG}(q^2)  \equiv G(q^2)
   =   1 + \sum_{n=1} g_n^{} a_s^n,
\eea
where  $N_{\!A} = 8$ in QCD.

The coefficients   $g_n$ of the renormalization of the absorptive part  in the  \MSb\ scheme corresponding to the self-energy of the Higgs boson  up to N$^4$LO are \cite{Herzog:2017dtz},
\begin{align}
\label{g1-4}
 g_1 \, =\,& 
    73  -\frac{14 }{3} n_f -\frac{46 }{3} L_q  \\ \nonumber 
 g_2 \, = \, & 3887.57 -629.982n_f+14.4283n_f^2 - \left[23 \left(73-\frac{14}{3}n_f\right)+\frac{464}{3}\right]L_q+\frac{529}{3}L_q^2
    \\ \nonumber 
 g_3 \, =\,&163394-49409.6n_f+2974.39n_f^2-34.4213n_f^3-\left[\frac{9769}{9}+\frac{580}{3} \left(73-\frac{14}{3}n_f\right)\right.\\ \nonumber
 &\left.-\frac{92}{3} \left(3887.57-629.982n_f+14.4283n_f^2\right)\right]L_q+\left[\frac{34684}{9}+\frac{1058}{3}\left(73-\frac{14}{3}n_f\right)\right]L_q^2-\frac{48668}{27}L_q^3\\ \nonumber
 g_4 \,=\,& 5.45154\times 10^6-2.81728\times 10^6n_f+318324.n_f^2-9921.43 n_f^3+64.359 n_f^4+\Biggl[-38609.3\\ \nonumber
 &-232\left(3887.57 - 629.982 n_f + 14.4283 n_f^2\right)-\frac{115}{3}\left(163394. - 49409.6 n_f + 2974.39 n_f^2 - 34.4213 n_f^3\right)\\ \nonumber
 &-\frac{68383}{54} \left(73-\frac{14}{3}n_f\right)\Biggr]L_q+\Biggl[\frac{3924805}{81}+\frac{57362}{9} \left(73-\frac{14}{3}n_f\right)+\frac{5290}{9} \left(3887.57 - 629.982n_f\right.\\ \nonumber
 & \left.+ 14.4283 n_f^2\right)\Biggr]L_q^2+\Biggl[\frac{5093212}{81}-\frac{121670}{27}\left(73-\frac{14 }{3}n_f\right)\Biggr]L_q^3+\frac{1399205 }{81}L_q^4,
\end{align}
where $L_q = ln \left( \frac{q^2}{\mu^2} \right)$.

The logarithm in equation \ref{g1-4} can also be written in terms of the pole mass of the top quark by defining,
\begin{equation}
\label{LqM}
L_q = T -ln \Bigl( \frac{\mu^2}{M_t^2} \Bigr),
\end{equation}
where $T\equiv ln \left( M_H^2 / M_t^2\right)$ and $M_t$ is the pole mass of the top quark, where we have set $q^2 = M_H^2$.
\section{Fixed-order-perturbation theory}
\label{sec3}
In the RGSPT,  predicting RG-accessible next-order coefficients becomes more effective if we express the perturbative expansion in terms of the running fermion mass \cite{Elias:2000iw}.  Therefore, we rewrite  the $H \rightarrow gg$ decay rate expansion in terms of the running top-quark mass.  For this purpose,  we use the relation between the running and pole mass \cite{Chetyrkin:1999qi,Melnikov:2000qh},
\begin{align}
    \frac{m(\mu)}{M} =&
  1 
  +\left[-1-\frac{4}{3}l_{\mu M}\right] x (\mu)\\ \nonumber
 & +\left[  -14.3444-\frac{445}{72}l_{\mu M}-\frac{19}{24} l_{\mu M}^2
  +\left(1.04137+\frac{13}{36} l_{\mu M} +\frac{1}{12}l_{\mu M}^2\right)n_f\right]x (\mu)^2\\ \nonumber
 & +\left[  -198.707-78.9409 l_{\mu M}-\frac{11779}{864}l_{\mu M}^2-\frac{475}{432}l_{\mu M}^3+\Biggl(26.9239+12.3257l_{\mu M}\Biggr.\right.\\ \nonumber
 &\left.\left.+\frac{911}{432}l_{\mu M}^2+\frac{11}{54}l_{\mu M}^3\right)n_f+\left(-0.652692-0.380301l_{\mu M}^2-\frac{1}{108}l_{\mu M}^3\right)n_f^2\right]x (\mu)^3,
\end{align}
where  $\lmM=\ln\mu^2/M^2$,   $x (\mu)= \frac{\alpha_s^{n_f} (\mu)}{\pi}$ and $n_f$ are number of active flavours.

Using above relation,  the logarithm $L_q$ in the equation \ref{LqM}  can be expressed in terms of $ln \frac{\mu^2}{m_t^2 (\mu)}$.  This can be done by writing the logarithm $ ln \Bigl( \frac{\mu^2}{M_t^2} \Bigr)$  as,
\begin{align}
ln \Bigl( \frac{\mu^2}{M_t^2} \Bigr) \,=&\    ln \frac{\mu^2}{m_t^2 (\mu)} +2 x^{(n_f)}(\mu)\Biggl[-1.33333 -ln \frac{\mu^2}{m_t^2 (\mu)}\Biggr]\\ \nonumber
&+2x^{(n_f)}(\mu)^2\Biggl[-12.5666+1.04137n_f+\Bigl(-5.51389                                                          +0.361111n_f\Bigr) ln \frac{\mu^2}{m_t^2 (\mu)}\\ \nonumber
&+\Bigl(-1.29167+0.0833333n_f\Bigr)\left(ln \frac{\mu^2}{m_t^2 (\mu)}\right)^2\Biggr]+2x^{(n_f)}(\mu)^3\Biggl[-173.452+25.2666n_f-0.652691 n_f^2\\ \nonumber
&+\Bigl(-70.3593+11.9597n_f\Bigr) ln \frac{\mu^2}{m_t^2 (\mu)}
+\Bigl(-14.4525 + 2.08102n_f - 0.380301 n_f^2\Bigr)\left (ln \frac{\mu^2}{m_t^2 (\mu)}\right)^2\\ \nonumber
&+\Bigl(-2.22454 + 0.287037 n_f - 0.00925926 n_f^2\Bigr)\left (ln \frac{\mu^2}{m_t^2 (\mu)}\right)^3\Biggr].
\end{align}
Thus, the decay width of $H \rightarrow gg$ acquires  the following  form in terms of the running top quark mass,
 \begin{equation}
\Gamma = \left[ \sqrt{2} G_F M_H^3 / 72\pi \right] x^2 (\mu)
S\left[ x(\mu), L(\mu) \right],
\label{Gam_higgs}
\end{equation}
where the  perturbative expansion $S[x(\mu), L(\mu)] $ in the FOPT  is written as,
\begin{equation}
\label{fopt}
S_{\rm FOPT}[x(\mu), L(\mu)] =
\sum_{n=0}^\infty \sum_{k=0}^n T_{n,k} x^n L^k.
\end{equation}

We work with $n_f = 5$ flavours in this work.  The coefficients of expansion $T_{n,k}$ in the  \MSb\ scheme are,

\begin{align}
T_{0,0}^{\overline{\text{MS}}} &= 1, ~ T_{1,0}^{\overline{\text{MS}}} =17.9167\, -3.83333\, T , ~
T_{1,1}^{\overline{\text{MS}}} =3.83333 , ~
T_{2,0}^{\overline{\text{MS}}} = 146.586 - 102.146\, T + 11.0208 \,T^2, \\ \nonumber
T_{2,1}^{\overline{\text{MS}}} &=100.188 - 22.0417\, T , ~T_{2,2}^{\overline{\text{MS}}} =11.0208,  ~T_{3,0}^{\overline{\text{MS}}} =123.647 - 1156.52\, T + 394.514\, T^2 - 28.1644\, T^3 \\ \nonumber
T_{3,1}^{\overline{\text{MS}}} &=1034.83 - 766.826\, T + 84.4931\, T^2, ~ T_{3,2}^{\overline{\text{MS}}}  =  376.545 - 84.4931\, T,~ T_{3,3}^{\overline{\text{MS}}}  = 28.1644, \\ \nonumber
T_{4,0}^{\overline{\text{MS}}} &= -11815.8 - 2506.94\, T + 5777.77\, T^2 - 1274.79\, T^3 + 67.4771\, T^4, \\ \nonumber
T_{4,1}^{\overline{\text{MS}}} &=45.6137 - 10430.6\, T + 3718.31\, T^2 - 269.908\, T^3 , ~ T_{4,2}^{\overline{\text{MS}}} =4609.02 - 3615.6\, T + 404.863\, T^2 , \\ \nonumber
T_{4,3}^{\overline{\text{MS}}} &= 1184.31 - 269.908\, T , \, T_{4,4}^{\overline{\text{MS}}} = 67.4771,
\label{higgs_coeffs}
\end{align}
and the logarithm  is $L (\mu) = ln (\mu^2/m_t^2 (\mu))$.

We can also write the FOPT expansion in terms of SI, OS, and miniMOM schemes.  For this purpose,  we write equation  \ref{g1-4}  in terms of the SI, OS, and miniMOM mass of the top quark,
\begin{equation}
\label{Lq}
L_q = T -ln \Bigl( \frac{\mu^2}{M_{t_{p}}^2} \Bigr),
\end{equation}
where $T\equiv ln \left( M_H^2 / M_{t_{p}}^2 \right)$ and $p$ stands for SI,  OS or miniMOM scheme.   In the SI scheme,  the  coefficients of  the FOPT expansion are,
\begin{align}
T_{0,0}^{\text{SI}}=& 1,\, T_{1,0}^{\text{SI}}= 17.9167 - 3.83333\, T,\, T_{1,1}^{\text{SI}}=3.83333,\,T_{2,0}^{\text{SI}}=156.808 - 102.146 \,T + 11.0208\, T^2,\, \\ \nonumber
T_{2,1}^{\text{SI}}= & 107.854 - 22.0417\, T,
T_{2,2}^{\text{SI}}=11.0208,\,T_{3,0}^{\text{SI}}=452.461-1215.3 T+394.514 T^2-28.1644 T^3,\\ \nonumber
T_{3,1}^{\text{SI}}=&1337.74 - 810.91\, T + 84.4931\, T^2,\,T_{3,2}^{\text{SI}}=427.337 - 84.4931 T,\,T_{3,3}^{\text{SI}}=28.1644,\\ \nonumber
T_{4,0}^{\text{SI}}=&-6502.1 - 4935.46 T + 6003.09 T^2 - 1274.79 T^3 + 67.4771 T^4,\\ \nonumber
T_{4,1}^{\text{SI}}=&6041.54 - 12666.6 T + 3887.29 T^2 - 269.908 T^3,\,T_{4,2}^{\text{SI}}=6947.57 - 3992.14\, T + 404.863\, T^2,\\ \nonumber
T_{4,3}^{\text{SI}}=&1400.62 - 269.908\, T,\,T_{4,4}^{\text{SI}}=67.4771.
\end{align}
In a similar manner,  we write the coefficients of expansion  in the OS scheme,
\begin{align}
T_{0,0}^{\text{OS}}=& 1,\, T_{1,0}^{\text{OS}}= 17.9167 - 3.83333\, T,\, T_{1,1}^{\text{OS}}=3.83333,\,T_{2,0}^{\text{OS}}=156.808 - 102.146 \,T + 11.0208\, T^2,\, \\ \nonumber
T_{2,1}^{\text{OS}}= & 107.854 - 22.0417\, T,
T_{2,2}^{\text{OS}}=11.0208,\,T_{3,0}^{\text{OS}}=467.684 - 1215.3\, T + 394.514\, T^2 - 28.1644\, T^3,\\ \nonumber
T_{3,1}^{\text{OS}}=&1337.74 - 810.91\, T + 84.4931\, T^2,\,T_{3,2}^{\text{OS}}=427.337 - 84.4931 T,\,T_{3,3}^{\text{OS}}=28.1644,\\ \nonumber
T_{4,0}^{\text{OS}}=&-6091.71 - 4993.81\, T + 6003.09 \,T^2 - 1274.79\, T^3 + 67.4771\, T^4,\\ \nonumber
T_{4,1}^{\text{OS}}=&6187.42 - 12666.6\, T + 3887.29\, T^2 - 269.908\, T^3,\,T_{4,2}^{\text{OS}}=6947.57 - 3992.14\, T + 404.863\, T^2,\\ \nonumber
T_{4,3}^{\text{OS}}=&1400.62 - 269.908\, T,\,T_{4,4}^{\text{OS}}=67.4771.
\end{align}
We also investigate the Higgs to gluon decay width in the  miniMOM version of the OS scheme \cite{miniMOM1,miniMOM2}.  In this scheme, the strong coupling is fixed by the knowledge of the gluon and ghost propagators \cite{miniMOM1,miniMOM2}.   The expansion coefficients for this scheme can be written using the expansion given in reference  \cite{Herzog:2017dtz},
\begin{align}
T_{0,0}^{\text{MM}} &= 1,\,T_{1,0}^{\text{MM}}=13.6528\, -3.83333 T,\,T_{1,1}^{\text{MM}}=3.83333,\,T_{2,0}^{\text{MM}}=46.9335-83.3368 T+11.0208 T^2,\, \\ \nonumber
T_{2,1}^{\text{MM} } &=89.0451\, -22.0417 T,\,
T_{2,2}^{\text{MM}}=11.0208,\,T_{3,0}^{\text{MM}}=-624.885 - 495.626 T + 333.353 T^2 - 28.1644 T^3,\\ \nonumber
T_{3,1}^{\text{MM}}&=569.388 - 710.471 T + 84.4931 T^2,\,
T_{3,2}^{\text{MM}}=388.058 - 84.4931 T,\,T_{3,3}^{\text{MM}}=28.1644,\\ \nonumber
T_{4,0}^{\text{MM}} &=-7041.01 + 5157.94 T + 2696.63 T^2 - 1100.39 T^3 + 67.4771 T^4,\\ \nonumber
T_{4,1}^{\text{MM}}& =-5111.88 - 6155.32 T + 3510.88 T^2 - 269.909 T^3, 
T_{4,2}^{\text{MM}} =3626.15 - 3615.73 T + 404.863 T^2,\,\\ \nonumber T_{4,3}^{\text{MM}}&=1226.21 - 269.909 T,
T_{4,4}^{\text{MM}} =67.4771.
\end{align}
where MM represents the miniMOM scheme.    The running of the strong coupling in the miniMOM is given by \cite{N4LOmMOM}, 
\bea
  \alpha_{s,\rm MM} & =  & \alpha_s
   + 0.67862 \alpha_s^2
   + 0.91231  \alpha_s^3
   + 1.5961  \alpha_s^4
   + 3.1629  \alpha_s^5
   + \mathcal{O}( \alpha_s^6). 
\eea

\section{Renormalization-group-summed  perturbation theory}
\label{sec4}
Now, we discuss the perturbative expansion of the Higgs to gluons decay rate in the RGSPT.    In the RGSPT,   the FOPT expansion of the function $S[x(\mu), L(\mu)] $ is equivalent to writing the following new expansion,
\begin{equation}
\label{dseries}
 S (x, L) =
\sum_{n=0}^\infty x^n S_n (u),
\end{equation}
where  $u=x L$, and the function  $S_n (u) $ is defined by,
\begin{equation}\label{Dn_def}
S_n (u) \equiv \sum_{k=n}^\infty  T_{k, k-n} u^{k-n}.
\end{equation}
The main feature of the RGSPT is the explicit all-orders summations of  all RG-accessible logarithms in the functions $S_n (u) $ \cite{Ahmady1, Ahmady2}.  Moreover, the functions $S_n (u) $  can be derived in  a closed analytical form \cite{Ahmady1, Ahmady2}. 

We present a  derivation of  the functions $S_n (u) $ in a closed analytical form using the RG invariance.  The  decay width of the $H \rightarrow gg$  decay  defined in equation \ref{Gam_higgs}  is scale independent order-by-order in perturbation theory,   and  satisfies the RG equation, 
\begin{equation}
\mu^2 \frac{\mathrm{d}}{\mathrm{d}\mu^2} \left\{    \Gamma_{H \rightarrow  gg} \right\} =
0.
\end{equation}

The above equation  can be written in the following form:
\begin{equation}%6.28
\left[ \left( 1 - 2\gamma_m (x) \right) \frac{\partial}{\partial L} +
\beta (x) \frac{\partial}{\partial x} + \frac{2\beta(x)}{x} \right] S (x,
L) = 0,
\label{higgs_RG}
\end{equation}
where the $\overline{{\rm MS}}$ $\beta$- and $\gamma_m$-functions are,
\begin{align}
\begin{split}
\beta(x) & = \mu^2 \frac{\mathrm{d}}{\mathrm{d}\mu^2} x(\mu) = -\left( \beta_0 x^2 + \beta_1 x^3 + \beta_2 x^4+\dots\right), 
\\
\mu^2 \frac{\mathrm{d}m}{\mathrm{d}\mu^2}  &=  m \gamma_m (x(\mu))
 =  -m (\gamma_0 x + \gamma_1 x^2 + \gamma_2 x^3 + \dots).
\end{split}
\label{beta_6f}
\end{align}
The coefficients of 5-loop QCD $\beta$ -function in the $\overline{{\rm MS}}$  scheme read \cite{ChBK-LL16},
\begin{align}
\beta_0\,=\,&2.75\, -0.166667 n_f,\nonumber \\
\beta_1\,=\,&6.375\, -0.791667n_f ,\nonumber \\
\beta_2\,=\,& 22.3203 -4.36892 \,n_f + 0.0940394\,n_f^2 ,\nonumber \\
\beta_3\,=\,&114.23 -27.1339 \,n_f   +1.58238 \,n_f^2  + 0.0058567 \,n_f^3\nonumber ,\\
\beta_4\,=\,& 524.558  -181.799 \,n_f  +17.156 \,n_f^2 -0.225857 \,n_f^3 -0.00179929 \,n_f^4,
\label{beta_nf}
\end{align}
and the coefficients of the $\gamma_m$-function are given as\cite{Baikov:2014qja},
\begin{align}
\gamma_0\,=\,&1,\nonumber \\
\gamma_1\,=\,&4.20833\, -0.138889 n_f ,\nonumber \\
\gamma_2\,=\,&   19.5156  -2.28412 \,n_f   -0.0270062 \,n_f^2 ,\nonumber \\
\gamma_3\,=\,& 98.9434  -19.1075 \,n_f   +0.276163 \,n_f^2  +0.00579322 \,n_f^3   \nonumber ,\\
\gamma_4\,=\,& 559.707-143.686 \,n_f+7.48238 \,n_f^2  +0.108318 \,n_f^3   -0.0000853589 \,n_f^4. 
\label{gamma_nf}
\end{align}
By substituting the expansion in equation \ref{dseries} into the equation \ref{higgs_RG},   we write the  following equation,
\bea
0= \left( 1 - 2\gamma_m (x) \right)  \sum_{n=0}^\infty
\sum_{k=0}^n k  T_{n,k}  x^n L^{k-1}\nonumber
+ \beta(x)  \sum_{n=0}^\infty
\sum_{k=0}^n n T_{n,k}  x^{n-1} L^{k} + \frac{2\beta(x)}{x}  \sum_{n=0}^\infty \sum_{k=0}^n T_{n,k} x^n L^k.
\eea
The following recursion formula is derived by  extracting the aggregate coefficient of $x^n L^{n-p}$  for $n \geq p$:
\begin{equation}\label{recursion}
0 = (n-p+1) T_{n, n-p+1}- \sum_{\ell = 0} ^{p-1}  (n - \ell + 1) \beta_\ell T_{n - \ell - 1, n - p} + 2 \sum_{\ell = 1} ^{p-1}  (n - p + 1) \gamma_{\ell-1} T_{n - \ell , n - p + 1}.
\end{equation}
After this, we multiply both sides of equation (\ref{recursion})  by $ u^{n-p}$  and sum from $n=p$ to
$\infty$.  This results in a  set of first-order linear differential equations for the functions defined in equation (\ref{Dn_def}),
\begin{equation}
0  =(1-\beta_0 u) \frac{\mathrm{d} S_{p-1}}{\mathrm{d}u} - u \sum_{\ell = 0}^{p-2} \beta_{\ell+1} \frac{\mathrm{d} S_{p-\ell - 2}}{\mathrm{d}u}
 +2 \sum_{\ell = 0}^{p-2} \gamma_\ell \frac{\mathrm{d} S_{p-\ell - 2}}{\mathrm{d}u} -\sum_{\ell = 0}^{p-1} (p - \ell + 1) \beta_{\ell} S_{p - \ell - 1}.
\end{equation}
By substituting $n = p - 1$, the above equation can be written as,
\beq\label{Dk_de}
(1-\beta_0 u)  \frac{\mathrm{d}S_n}{\mathrm{d}u}  - u \sum_{\ell = 0}^{n-1} \beta_{\ell+1} \frac{\mathrm{d} S_{n-\ell -1}}{\mathrm{d}u}+2 \sum_{\ell = 0}^{n-1} \gamma_\ell \frac{\mathrm{d} S_{n-\ell -1}}{\mathrm{d}u} - \sum_{\ell = 0}^{n} (n - \ell + 2) \beta_{\ell} S_{n - \ell } =0,
\eeq
where $n, \ell \ge 0$ with the boundary condition $S_n (0) = T_{n,0}$.

We can now solve the system of equations (\ref{Dk_de})   iteratively in an analytical closed-form.   The solutions for $n=0,1,2,3$ are,
\begin{align}\label{D12}
S_0 (u) \, =\,&\frac{T_{00}}{w_1^2} ,\nonumber \\
S_1 (u) \,=\,&\frac{1}{\beta_0 w_1^3}\Bigl[-2 T_{00} (\beta_1-2 \beta_0 \gamma_0) \log w_1+\beta_0 T_{00}\Bigr],\nonumber \\
S_2(u) \,=\,&\frac{1}{\beta_0^2w_1^4}\Bigl[\beta_0 \left(-4 \beta_0^2 \gamma_1 T_{00} u+\beta_0 (2 T_{00} u (2 \beta_1 \gamma_0+\beta_2)+T_{20})-2 \beta_1^2 T_{00}
   u\right)-(\beta_1-2 \beta_0 \gamma_0) \log w_1 \nonumber \\ &(-4 \beta_0 \gamma_0 T_{00}+3 \beta_0 T_{00}+2 \beta_1 T_{00})+3 T_{00}
   (\beta_1-2 \beta_0 \gamma_0)^2 \log ^2w_1\Bigr],\nonumber \\
S_3(u) \,=\,&\frac{1}{\beta_0^3w_1^5}\Bigl[-\beta_0^2 \Bigl(-2 \beta_0^3 \gamma_2 T_{00} u^2+\beta_0^2 u (2 \beta_1 \gamma_1 T_{00} u+2 \beta_2 \gamma_0 T_{00} u+\beta_3
   T_{00} u-8 \gamma_0 \gamma_1 T_{00}+6 \gamma_1 T_{20}+4 \gamma_2 T_{00})\nonumber \\ &-\beta_0 \left(u \left(2 \beta_1^2 \gamma_0 T_{00} u+2
   \beta_1 \beta_2 T_{00} u-8 \beta_1 \gamma_0^2 T_{00}+6 \beta_1 \gamma_0 T_{20}+3 \beta_2 T_{20}+2 \beta_3
   T_{00}\right)+T_{30}\right)\nonumber \\ &+\beta_1 u (\beta_1 (\beta_1 T_{00} u-4 \gamma_0 T_{00}+3 T_{20})+2 \beta_2 T_{00})\Bigr)-\beta_0
   (\beta_1-2 \beta_0 \gamma_0) \log w_1 \Bigl(4 \beta_1 \gamma_0 T_{00} (1-3 \beta_0 u)\nonumber \\ &-2 \beta_2 (3 \beta_0 T_{00}
   u+T_{00})+2 \beta_0 \left(6 \beta_0 \gamma_1 T_{00} u-4 \gamma_0^2 T_{00}+3 \gamma_0 T_{20}+2 \gamma_1 T_{00}-2 T_{20}\Bigr)+6
   \beta_1^2 T_{00} u-3 \beta_1 T_{20}\right)\nonumber \\ &-(\beta_1-2 \beta_0 \gamma_0)^2 \log ^2w_1 (-14 \beta_0 \gamma_0 T_{00}+6 \beta_0
   T_{20}+7 \beta_1 T_{00})+4 T_{00} (\beta_1-2 \beta_0 \gamma_0)^3 \log ^3w_1\Bigr],
\end{align}
where $w_1=1-\beta_0u$.

 The new RGS expansions now can be written as,
\begin{align}
S^{NLO}_{RGSPT}\,=&\,S_0(xL)+xS_1(xL), \\ \nonumber 
S^{N^2LO}_{RGSPT}\,=&\,S_0(xL)+xS_1(xL)+x^2S_2(xL),\\ \nonumber 
S^{N^3LO}_{RGSPT}\,=&\,S_0(xL)+xS_1(xL)+x^2S_2(xL)+x^3S_3(xL),\\ \nonumber 
S^{N^4LO}_{RGSPT}\,=&\,S_0(xL)+xS_1(xL)+x^2S_2(xL)+x^3S_3(xL)+x^4S_4(xL).
\end{align}
In the SI,  OS, and miniMOM schemes,   mass appearing in the logarithms does not depend on the RG scale $\mu$.  Therefore,  the closed-form analytic expressions of the functions $S_n (u) $ in these schemes are obtained by solving equation \ref{Dk_de} after substituting the coefficient of the anomalous  $\gamma$ functions $\gamma_i$ to be zero.

\section{Scale and scheme dependence in the FOPT}
\label{SSDF}
We now investigate the perturbative behaviour of the Higgs to gluons decay width by studying the  scale  dependence in the FOPT  in   the $\overline{\text{MS}}$,  SI,  OS, and miniMOM schemes.    Our numerical inputs are given in table \ref{tab1}.  Moreover, for running the strong coupling and quark masses,  we use the Mathematica package ``RunDec" \cite{Chetyrkin:2000yt}.
   \begin{table}[H]
\begin{center}
   \begin{tabular}{|c|c|} \hline
Parameter & Value  \\  \hline
  $M_H$  &  $125.25 \pm0.17$   \cite{Zyla:2021}  \\  \hline
$G_{F}$ & $1.1663787 \times 10^{-5}$ \,$~{\rm GeV}$  \cite{Zyla:2021}\\   \hline
  $\alpha_{s}[M_{Z}]$ & $0.1179 \pm 0.0009$ \cite{Zyla:2021}   \\ \hline
  $M_{t_{OS}} $  &$ 173$ $~{\rm GeV}$  \cite{Zyla:2021}  \\ \hline
  $M_{t_{SI}} $          &$ 164$ $~{\rm GeV}$  \cite{Zyla:2021} \\ \hline   
\end{tabular} 
\end{center}
\caption{The numerical values of masses and couplings used in this work.}
   \label{tab1}
   \end{table}

The Higgs to gluons decay width is normalized by the first term in the expansion, i.e. $\Gamma_0$ given by,
\beq
  \Gamma_0 = G_F M_H^3 / (36 \pi^3 \sqrt{2})  (\alpha_s( M_H^2))^2=0.00018378,
\eeq
where  $\alpha_s( M_H^2) = 0.112602$ is calculated using the Mathematica package ``RunDec" \cite{Chetyrkin:2000yt}.  We use the above value of $ \Gamma_0$ in all our predictions.

\begin{figure}[H]
    \centering
    \subfigure[]{\includegraphics[width=0.45\textwidth]{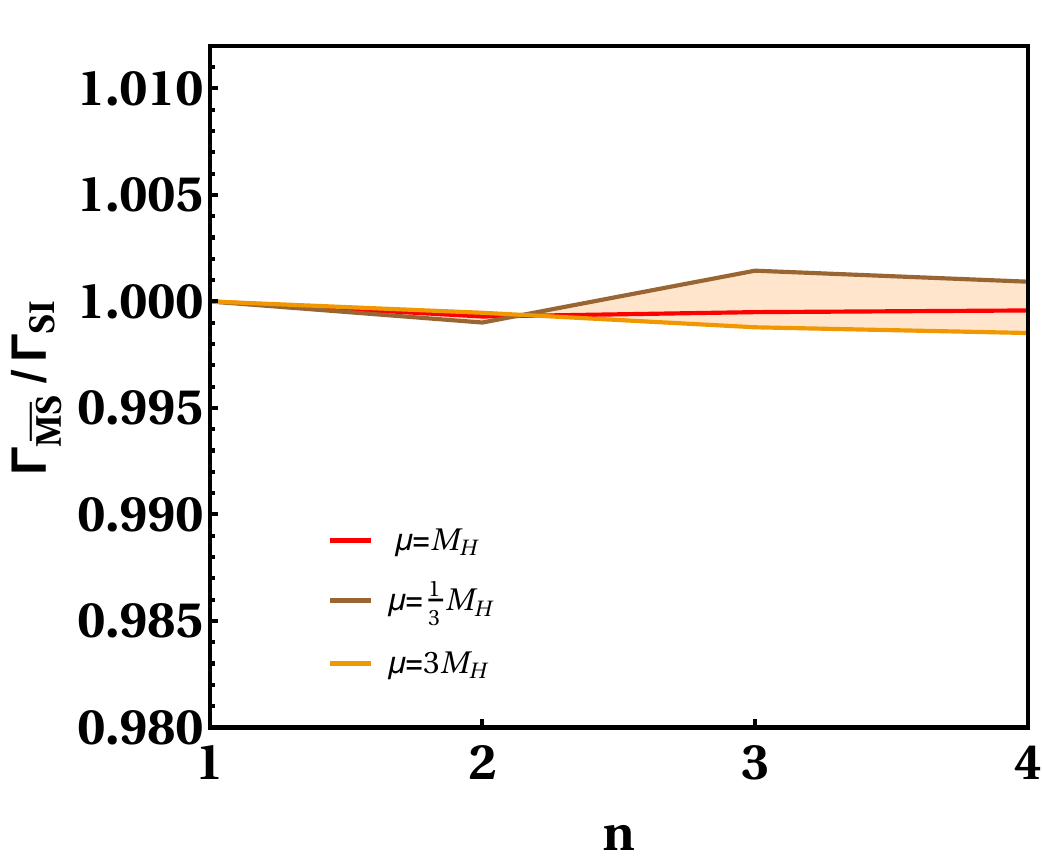}} 
    \subfigure[]{\includegraphics[width=0.45\textwidth]{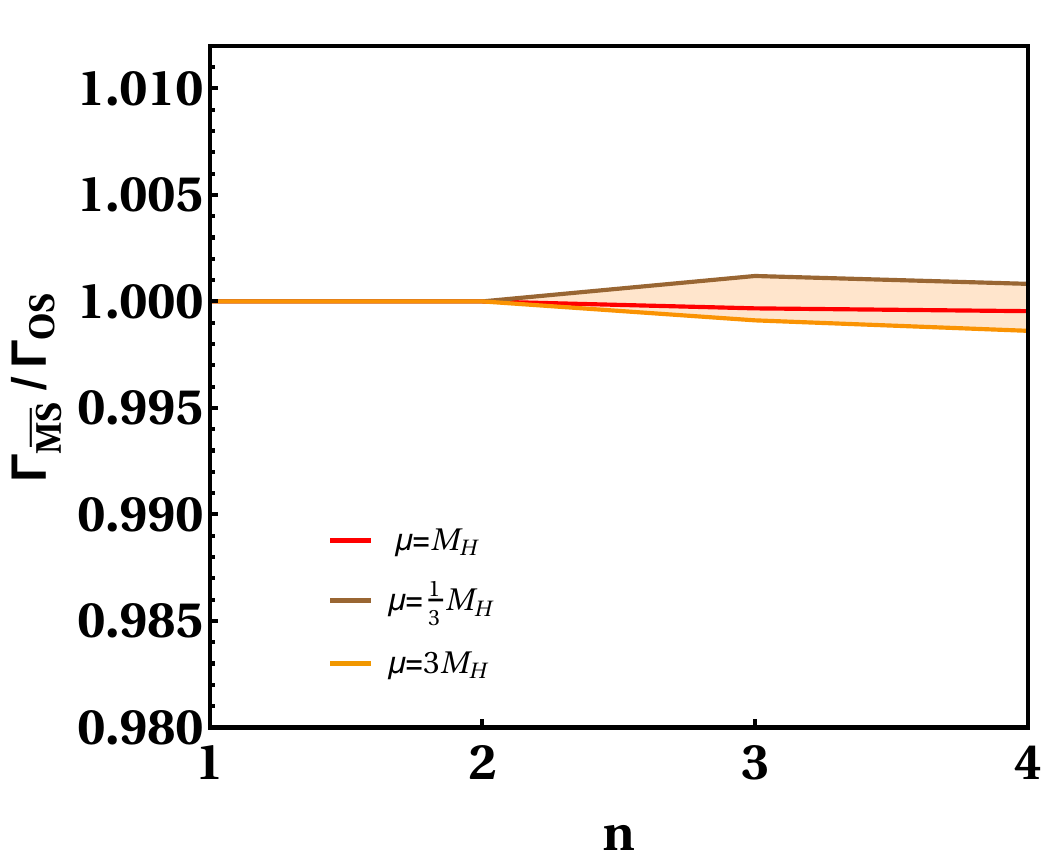}} 
    \subfigure[]{\includegraphics[width=0.45\textwidth]{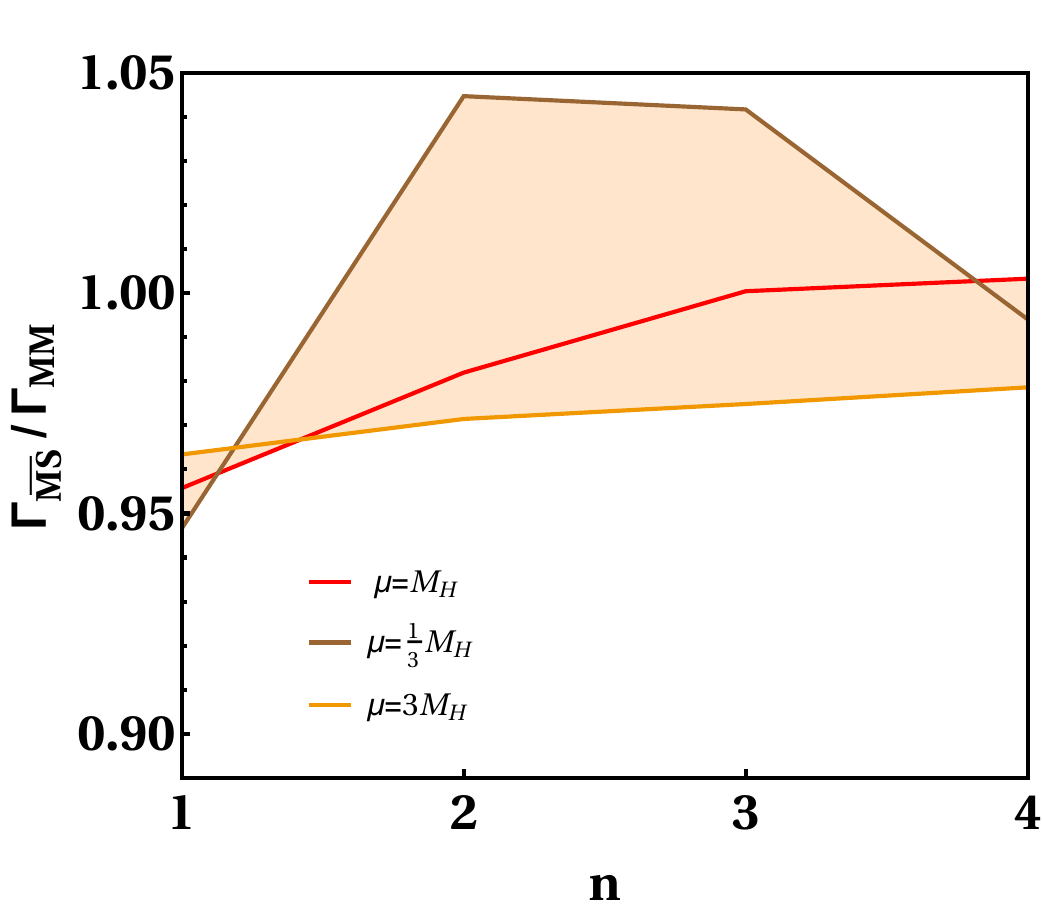}}
    \caption{The variation of  $\Gamma_{\overline{\text{MS}}}/\Gamma_{\text{scheme}}$   at RG scales $\mu= \frac{1}{3}M_H, M_H$, and $ 3 M_H$  in the  FOPT in the (a)SI (b)OS and (c)miniMOM schemes up to order $n=4$.}
    \label{scale_var_fopt}
\end{figure}

We show the RG scale dependence of  the ratio of the Higgs to gluons decay width in two different RG schemes in the FOPT in figure  \ref{scale_var_fopt}.   To compare the predictions of different RG schemes, we have chosen the  $\overline{\text{MS}} $ scheme as the reference scheme.   The behaviour of the OS scheme is closer to that of the $\overline{\text{MS}} $ scheme due the same mass of the top quark used in both schemes.

In  the $\overline{\text{MS}} $ scheme,    the contribution   of the $\rm N^4LO$ correction  (defined by $\Gamma_{\rm N^4LO} \times 100/\Gamma$) to the $\Gamma (H \rightarrow gg)$ decay width  at the renormalization scale   $\mu= M_H $  for the on shell top quark mass   is $-0.6\%$.    We now vary the scale for a  range of $\mu= \frac{1}{3} M_H$ to $\mu= 3 M_H$,  and find that the contribution  of the $\rm N^4LO$ corrections to the $\Gamma (H \rightarrow gg)$ in the  $\overline{\text{MS}}$ scheme is $0.2\%$ and $2\%$,  respectively.   The decay width in this range is  $\Gamma (H \rightarrow gg) =1.836  \Gamma_0$ at the RG scale $\mu= \frac{1}{3} M_H$,   $\Gamma (H \rightarrow gg) =1.842 \Gamma_0$ at the RG scale $\mu= M_H$,   and $\Gamma (H \rightarrow gg) =1.838\Gamma_0$ at the RG scale $\mu=3 M_H$.   The $\rm N^4LO$  contribution in the SI and OS schemes is approximately identical to that of the $\overline{\text{MS}} $ scheme.   The decay width in this range is  $\Gamma (H \rightarrow gg) =1.834  \Gamma_0$ at the RG scale $\mu= \frac{1}{3} M_H$,   $\Gamma (H \rightarrow gg) =1.843 \Gamma_0$ at the RG scale $\mu= M_H$,   and $\Gamma (H \rightarrow gg) =1.841 \Gamma_0$ at the RG scale $\mu=3 M_H$ in the SI and OS schemes.

On the other hand,  the $\rm N^4LO$  contribution  in the miniMOM scheme  is      $-1.06 \%$ at  RG scale $\mu= M_H $.    At  RG scale $\mu= M_H/3$,  the  $\rm N^4LO$ correction in the miniMOM scheme is $5.55 \%$ which is quite large. The $\rm N^4LO$  contribution at RG scale $\mu= 3 M_H$ is $0.68\%$. The decay width in this range is  $\Gamma (H \rightarrow gg) =1.847  \Gamma_0$ at the RG scale $\mu= \frac{1}{3} M_H$,   $\Gamma (H \rightarrow gg) =1.836 \Gamma_0$ at the RG scale $\mu= M_H$,   and $\Gamma (H \rightarrow gg) =1.879 \Gamma_0$ at the RG scale $\mu=3 M_H$ in this scheme.    We observe that the miniMOM scheme is quite sensitive to the scale variation,  and relatively away from the predictions of the  $\overline{\text{MS}}$ scheme.

\section{Scale and scheme dependence in the RGSPT}
\label{SSDR}
\begin{figure}[H]
    \centering
    \subfigure[]{\includegraphics[width=0.45\textwidth]{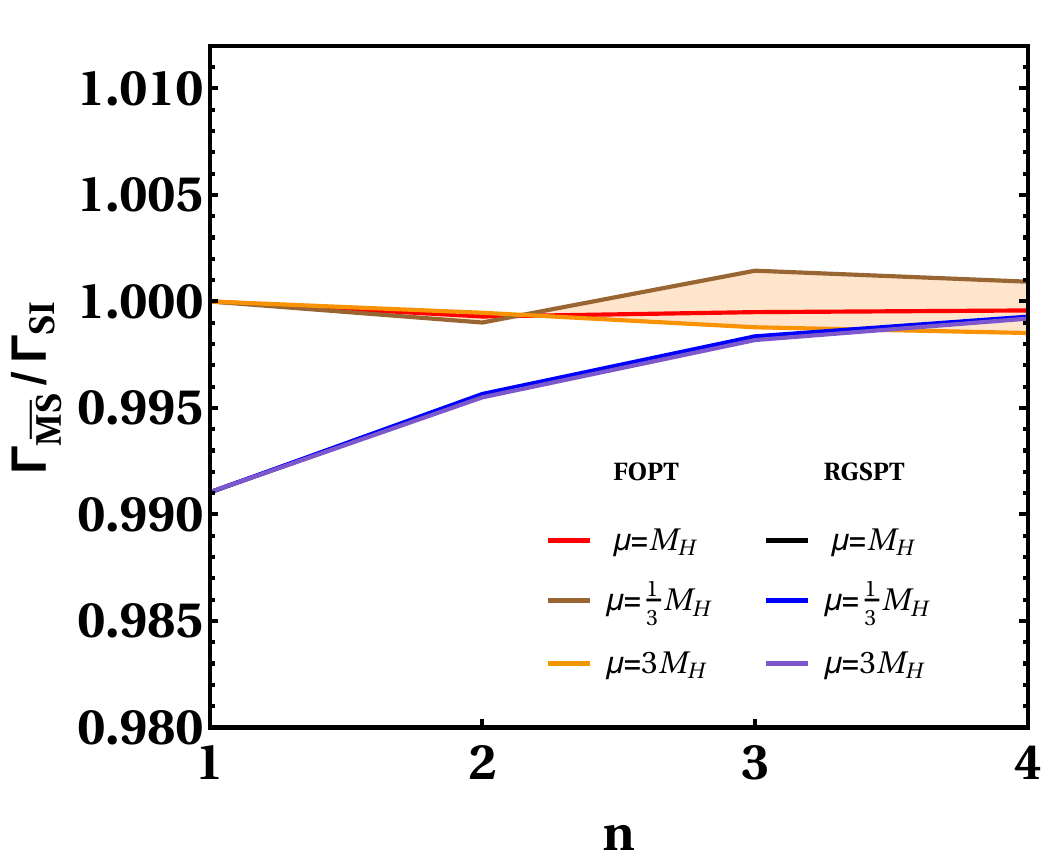}} 
    \subfigure[]{\includegraphics[width=0.45\textwidth]{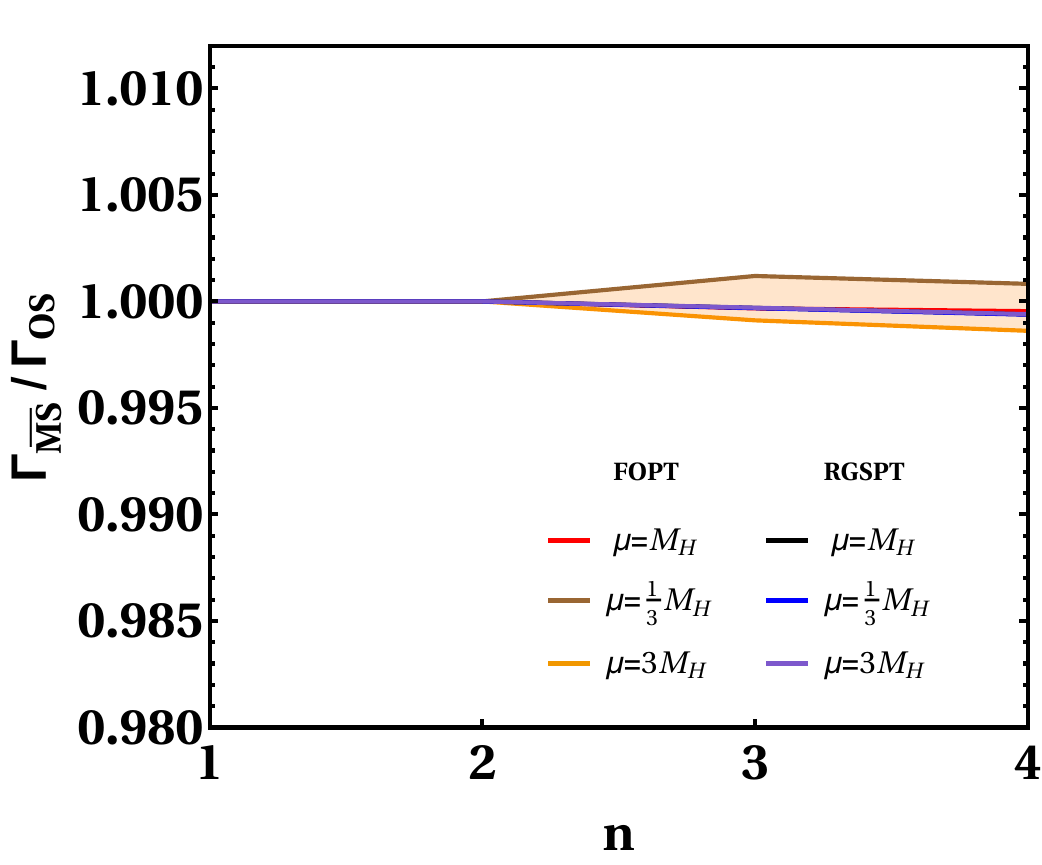}} 
    \subfigure[]{\includegraphics[width=0.45\textwidth]{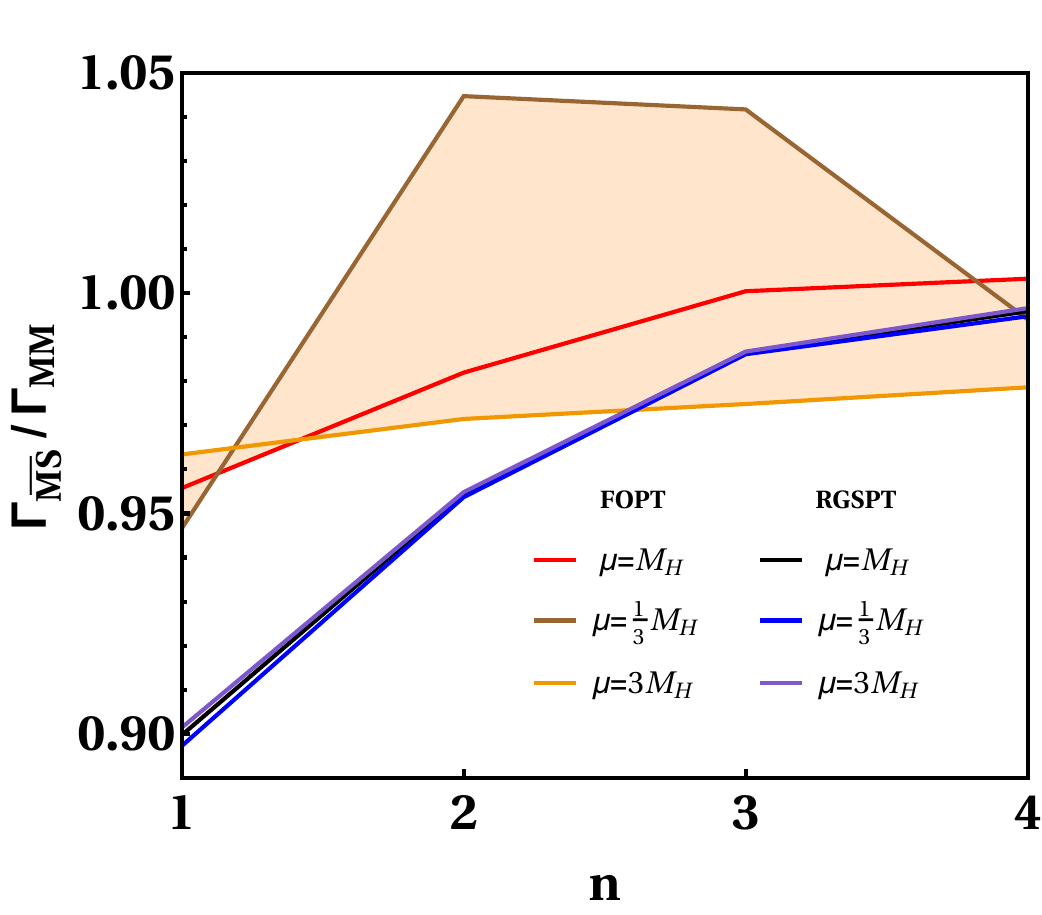}}
    \caption{The variation of    $\Gamma_{\overline{\text{MS}}}/\Gamma_{\text{scheme}}$    at RG scales $\mu= \frac{1}{3}M_H, M_H$, and $ 3M_H$  in the  FOPT and RGSPT   in the  (a)SI (b)OS and (c)miniMOM schemes up to order  $n=4$.}
    \label{scale_var1}
\end{figure}

In this section,  we discuss  the behaviour of the RGSPT expansions of  the Higgs to gluons decay width   in the  $\overline{\text{MS}}$,  SI,  OS, and miniMOM  schemes.      The variation  of  the ratio of the Higgs to gluons decay width in two different RG schemes at different orders  is  shown in figure \ref{scale_var1}.   There is a relatively small difference between the  FOPT and RGSPT predictions of the $\Gamma_{\overline{\text{MS}}}/\Gamma_{\text{SI}}$ upto order $n=3$ due to the different numerical values of the top-quark mass used in these schemes.

We begin our discussion by studying  the behaviour of the  RGSPT expansions in the $\overline{\text{MS}}$ scheme at  the RG scale  $\mu= M_H $  using the on shell top quark mass.   The contribution of the $\rm N^4LO$ correction to the  $\Gamma (H \rightarrow gg)$ decay width   at  the RG scale  $\mu= M_H $ is found to be $-0.06\%$ in the  $\overline{\text{MS}}$ scheme.    At  the RG scale  $\mu= \frac{1}{3} M_H$,  this contribution becomes $-0.18\%$ and changes to $0.04\%$ at  the RG scale $\mu= 3 M_H$.  The decay width at the RG scales $\mu=\frac{1}{3} M_H$, $\mu= M_H$, and $\mu= 3M_H$    is  $\Gamma (H \rightarrow gg) =1.845  \Gamma_0$ . This actually shows that the RGSPT expansion is considerably less sensitive to  the RG scale $\mu$.    We observe  that at the RG scale  $\mu= M_H $,   the $\rm N^4LO$ correction  in the SI and OS schemes  is  $-0.16\%$ and $-0.03\%$, respectively.  This correction becomes  $-0.28\%$ in the SI scheme and   $-0.15\%$ in the OS scheme at the RG scale $\mu= \frac{1}{3} M_H$,  and is $-0.06\%$ and $0.07\%$, respectively,  at the RG scale  $\mu= 3 M_H$.

Now we turn our attention towards the miniMOM scheme,  and discuss our results in this scheme for the RGSPT expansions.  We have already observed that in the FOPT,  this scheme is relatively more sensitive to the RG scale $\mu$.  The $\rm N^4LO$  contribution  in this scheme at  $\mu= M_H $ is  $-1.01\%$,  which is  large as compared to other used schemes.  The $\rm N^4LO$  contribution at RG scales $\mu= \frac{1}{3} M_H$   and    $\mu= 3 M_H$ is $-1.05\%$   and $-0.96\%$ respectively. The important observation is that this  $\rm N^4LO$  contribution remains stable relative  to the RG scale  in the RGSPT expansion  even at  $\mu= \frac{1}{3} M_H$   and    $\mu= 3 M_H$. The decay width in this range is  $\Gamma (H \rightarrow gg) =1.855  \Gamma_0$ at the RG scale $\mu= \frac{1}{3} M_H$,   $\Gamma (H \rightarrow gg) =1.853 \Gamma_0$ at the RG scale $\mu= M_H$,   and $\Gamma (H \rightarrow gg) =1.851 \Gamma_0$ at the RG scale $\mu=3 M_H$ in this scheme.

\section{Asymptotic Pad\'e approximant improved  $H \rightarrow gg$ decay rate}
\label{APAP}
The Pad\'e approximant is a nonlinear method of summing series that can be considered as an approximate analytic continuation \cite{baker}.  It is widely applied to determine higher-order terms in a number of field-theoretical perturbative expansions, including $\beta$- and $\gamma$-functions of QCD at four- and five-loops \cite{Samuel:1992xd,Samuel:1994jv,Samuel:1995jc,Ellis:1996zn,Ellis:1995jv,Ellis:1997sb,Chetyrkin:1997dh,vanRitbergen:1997va,Elias:1998bi}.

We begin by considering a generic perturbative expansion of the form,
\begin{equation}
\label{pert_exp}
S \equiv 1 + R_1 x + R_2 x^2 + R_3 x^3 +  R_4 x^4  + \cdots,
\end{equation}
where the coefficients $\{R_1, R_2, R_3, R_4 \}$ are known and the coefficients $\{  R_5, \cdots\}$ are unknown.

The Pad\'e approximant  to a generic perturbative expansion is denoted by,
\begin{eqnarray}
S_{[N|M]} & \equiv & \frac{1+a_1 x + a_2 x^2 + \cdots +  a_N x^N}{1 + b_1 x +
b_2 x^2 + \cdots + b_M x^M} \nonumber \\
& = & 1 + R_1 x + R_2 x^2 + R_3 x^3 + R_4 x^4 + \cdots + R_{N+M+1} \; x^{N+M+1} +\cdots.
\end{eqnarray}
The asymptotic error in the Pad\'e approximant prediction is given by\cite{Ellis:1996zn},
\begin{equation}
\label{error_for}
\delta_{N+M+1} =\frac{R_{N+M+1}^{Pad\acute{e}} - R_{N+M+1}} {R_{N+M+1}} = - \frac{M!
A^M}{[N+M+aM+b]^M}=\delta_i,
\end{equation}
where $R_{N+M+1}^{Pad\acute{e}}$ is the $[N|M]$ Pad\'e approximant prediction and   $R_{N+M+1}$ is the exact value of this coefficient.  The free parameters $\{A,a,b\}$ are chosen to produce the best results  \cite{Ellis:1996zn}.  In this section,  we  choose $a=-b=10^6$ which provide the best predictions.  Moreover,  for such large values of $a$ and $b$,  the error $\delta_i$ becomes very small.

If the coefficient  $R_1$   is known,  we can choose $N=0$ and $M=1$ and write the $[0|1]$ Pad\'e approximant as,
\begin{equation}
S_{[0|1]} = \frac{1}{1+b_1 x} = 1 - b_1 x + b_1^2 x^2 + \cdots = 1 + R_1 x
+ R_2^{Pad\acute{e}} x^2 + \cdots,
\end{equation}
where  $R_2^{Pad\acute{e}} = R_1^2$ is the Pad\'e approximant  prediction for the  next unknown higher-order term in the generic perturbative series in equation  \ref{pert_exp}.

The asymptotic error formula in equation \ref{error_for} provides the following equation for the $[0|1]$ Pad\'e approximant,
\begin{equation}
\frac{R_2^{Pad\acute{e}} - R_2}{R_2} = \frac{R_1^2 - R_2}{R_2} = \frac{
-A}{1+(a+b)},
\end{equation}
which gives the value of $A$ as,
\be
A =(1+a+b) \Bigl(1- \frac{R_1^2}{R_2}\Bigr).
\ee
We can determine the $[2|2]$ Pad\'e approximant  given the knowledge of only $R_1$,  $R_2$,  $R_3$ and   $R_4$. It is given by,
\begin{eqnarray}
S_{[2|2]} = \frac{1 + a_1 x + a_2 x}{1 + b_1 x + b_2 x} & = & 1 + \left(a_1 - b_1\right) x +
 \left(a_2 - a_1 b_1 + b_1^2 - b_2\right) x^2 \nonumber \\
&+&  \left(-a_2 b_1 - b_1^3 + a_1 (b_1^2 - b_2) + 2 b_1 b_2 \right) x^3  \nonumber \\
&+&  \left(-a_1 b_1^3 + b_1^4 + a_2 (b_1^2 - b_2) + 2 a_1 b_1 b_2 - 3 b_1^2 b_2 + b_2^2 \right) x^4  \nonumber \\
&+& \left(     -b_1^5 + 4 b_1^3 b_2 - 3 b_1 b_2^2 - a_2 (b_1^3 - 2 b_1 b_2) +  a_1 (b_1^4 - 3 b_1^2 b_2 + b_2^2)         \right) x^5 + \cdots \nonumber \\
& = & 1 + R_1 x + R_2 x^2  + R_3 x^3 + R_4 x^4  + R_5^{Pad\acute{e}} x^5 + \cdots.
\end{eqnarray}
The  $[2|2]$ Pad\'e approximant predicts, 
\begin{align}
R_5^{Pad\acute{e}} = \frac{R_3^3 - 2 R_2 R_3 R_4 + R_1 R_4^2}{-R_2^2 + R_1 R_3}.
\end{align}
The asymptotic error can be written as,
\begin{equation}
\label{delta_i}
\frac{R_5^{Pad\acute{e}} - R_5}{R_5}  = \frac{-2 A^2}{(4+2 a+b)^2} \equiv \delta_5.
\end{equation}
This error allows us to estimate an improved value of the  true value $R_5$, which is referred to as asymptotic Pad\'e approximant prediction (APAP) in literature \cite{Ellis:1996zn}.  Thus,  our APAP estimate of the   true value $R_5$ is,
\begin{eqnarray}
R_5 & = & \frac{(-R_3^3 + 2 R_2 R_3 R_4 - R_1 R_4^2)}{(1 + \delta_5)(R_2^2 - R_1 R_3)} \nonumber \\
 &=& \frac{8 R_2^2 (R_3^3 - 2 R_2 R_3 R_4 + R_1 R_4^2)}{(R_1^4 - 2 R_1^2 R_2 - 7 R_2^2) (R_2^2 - R_1 R_3)}.
 \label{R5_P}
\end{eqnarray}

Now, we apply the APAP formalism to predict $\rm N^5LO$  term  of the $\Gamma (H \rightarrow gg)$ decay width in the FOPT.  For the sake of clarity and reliability,  we present our analysis in the $\overline{\text{MS}} $,  SI, OS   and miniMOM schemes.  The $\Gamma (H \rightarrow gg)$ decay width is given in equation \ref{Gam_higgs} where  the  perturbative expansion $S[x(\mu), L(\mu)] $ now can be written as,
\begin{equation}
\label{fopt}
S_{\rm FOPT}[x(\mu), L(\mu)] =
\sum_{n=0}^\infty \sum_{k=0}^n T_{n,k} x^n L^k = \sum_{n=0}^\infty R_n x ^n,
\end{equation}
where
\begin{equation}
\label{fopt}
R_n [L(\mu)] =
 \sum_{k=0}^n T_{n,k} L^k.
\end{equation}
such that 
 \begin{equation}
 R_5[L]= T_{5,0}+ T_{5,1} L+T_{5,2}L^2+T_{5,3}L^3+T_{5,4}L^4+T_{5,5}L^5,
 \label{R5}
 \end{equation}
where $T_{5,0}, T_{5,1},T_{5,2},T_{5,3},T_{5,4}$ and $T_{5,5}$ are the unknown higher-order coefficients.  The coefficients  $T_{5,1}- T_{5,5}$ can be predicted with the help of the  RG invariance of $\Gamma (H \rightarrow gg)$ decay width.  The coefficient $T_{5,0}$ is not available through the  RG invariance.  

Thus,  we use equation \ref{higgs_RG} obtained from the  RG invariance of $\Gamma (H \rightarrow gg)$ decay width to predict the coefficients  $T_{5,1}- T_{5,5}$.  This means that the perturbative validity of equation \ref{higgs_RG}  to order   $x^5 L^5$ can determine the six-loop coefficients provided the $x^5$ contribution of equation  \ref{higgs_RG} vanishes.  Thus,  in the $\overline{\text{MS}}$ scheme  we obtain,

\begin{align}
T_{5,1}^{\overline{\text{MS}}}\,&=\, -0.251403 n_f^5+52.4477 n_f^4-2974.99 n_f^3+64964.7 n_f^2-569877. n_f+1.58072\times 10^6 , \nonumber \\
T_{5,2}^{\overline{\text{MS}}}\,&=\, 2.23573 n_f^4-306.398 n_f^3+11315.1 n_f^2-148924. n_f+600692, \nonumber \\
T_{5,3}^{\overline{\text{MS}}}\,&=\,  -0.0357832 n_f^4-7.47604 n_f^3+752.153 n_f^2-16703.8 n_f+104439, \nonumber \\
T_{5,4}^{\overline{\text{MS}}}\,&=\, -0.00462963 n_f^4+0.319155 n_f^3+12.6003 n_f^2-784.131 n_f+8181.75, \nonumber \\
T_{5,5}^{\overline{\text{MS}}}\,&=\,  222.674\, -13.4954 n_f. 
\label{RG_coeff_ms}
\end{align}

In the SI scheme, the coefficients are as follows,

\begin{align}
T_{5,1}^{\text{SI}}\,&=\,   -0.251403 n_f^5+51.897 n_f^4-2904.65 n_f^3+62487.4 n_f^2-538466 n_f+1.45824\times 10^6 ,\nonumber \\
T_{5,2}^{\text{SI}}\,&=\, 2.23573 n_f^4-302.321 n_f^3+11022.4 n_f^2-143316. n_f+569814, \nonumber \\
T_{5,3}^{\text{SI}}\,&=\,   -0.0357832 n_f^4-7.30602 n_f^3+737.153 n_f^2-16307.8 n_f+101326, \nonumber \\
T_{5,4}^{\text{SI}}\,&=\,  -0.00462963 n_f^4+0.319155 n_f^3+12.6003 n_f^2-776.921 n_f+8062.79,\nonumber \\
T_{5,5}^{\text{SI}}\,&=\,  222.674\, -13.4954 n_f.
\label{RG_coeff_si}
\end{align}

The coefficients obtained in the OS scheme are,

\begin{align}
T_{5,1}^{\text{OS}}\,&=\,  -0.251403 n_f^5+52.4682 n_f^4-2966.69 n_f^3+64629.9 n_f^2-566720. n_f+1.57759\times 10^6, \nonumber \\
T_{5,2}^{\text{OS}}\,&=\,  2.23573 n_f^4-305.229 n_f^3+11256.9 n_f^2-148434. n_f+602130, \nonumber \\
T_{5,3}^{\text{OS}}\,&=\,  -0.0357832 n_f^4-7.29085 n_f^3+744.961 n_f^2-16652.6 n_f+104721, \nonumber \\
T_{5,4}^{\text{OS}}\,&=\,  -0.00462963 n_f^4+0.319155 n_f^3+12.6003 n_f^2-784.131 n_f+8181.75,\nonumber \\
T_{5,5}^{\text{OS}}\,&=\,  222.674\, -13.4954 n_f,
\label{RG_coeff_os}
\end{align}

whereas the coefficients obtained in the miniMOM scheme are,

\begin{align}
T_{5,1}^{\text{MM}}\,&=\,   -0.00359858 n_f^4-0.168327 n_f^3+150.512 n_f^2+6047.03 n_f-136552, \nonumber \\
T_{5,2}^{\text{MM}}\,&=\, 0.033676 n_f^3+28.524 n_f^2-3364.45 n_f+25534.1, \nonumber \\
T_{5,3}^{\text{MM}}\,&=\,   1.38186 n_f^2-2691.8 n_f+38749.9, \nonumber \\
T_{5,4}^{\text{MM}}\,&=\, 6001.76\, -378.012 n_f,\nonumber \\
T_{5,5}^{\text{MM}}\,&=\,  222.675\, -13.4954 n_f.
\label{RG_coeff_MM}
\end{align}

Now, we use the  APAP method to predict the six-loop coefficients $\{T_{5,0}-T_{5,5}\}$ in the  $\overline{\text{MS}} $,  SI,  OS,  and miniMOM schemes.  The reliability of these APAP predicted coefficients will be estimated by calculating the uncertainty in their predictions against the RGE predicted $\{T_{5,1}-T_{5,5}\}$  coefficients.  For this purpose,  we define the moments of $R_5(w)$,  where  $w = m_t^2 (\mu) / \mu^2\ ;  (L = -ln (w) )$,  over the perturbative region $0<w\le 1$ in the following manner,
\begin{equation}%3.9
N_k \equiv (k+2) \int_0^1 dw ~\; w^{k+1} R_5 (w).
\label{mom_int}
\end{equation}
We substitute equation  \ref{R5} into equation  \ref{mom_int}, which results in a set  of equations for moments given by,
\begin{align}
N_{-1}\,&=\,T_{5,0}+T_{5,1}+2 \Bigl(T_{5,2}+3T_{5,3} +12 T_{5,4}+60 T_{5,5}\Bigr), \nonumber \\
N_{0}\,&=\,\frac{1}{4}\Bigl (4 T_{5,0}+2 T_{5,1}+2 T_{5,2}+3 T_{5,3}+6 T_{5,4}+15 T_{5,5}\Bigr) ,\nonumber \\
N_{1}\,&=\,\frac{1}{81}\Bigl (81 T_{5,0}+27 T_{5,1}+18 T_{5,2}+18 T_{5,3}+24 T_{5,4}+40 T_{5,5}\Bigr), \nonumber \\
N_{2}\,&=\,\frac{1}{128}\Bigl (128 T_{5,0}+32 T_{5,1}+16  T_{5,2}+12 T_{5,3}+12 T_{5,4}+15 T_{5,5}\Bigr) ,\nonumber \\
N_{3}\,&=\,\frac{1}{625}\Bigl (625  T_{5,0}+125  T_{5,1}+50  T_{5,2}+30  T_{5,3}+24  T_{5,4}+24  T_{5,5}\Bigr) ,\nonumber \\
N_{4}\,&=\,\frac{1}{324}\Bigl (324 T_{5,0}+54 T_{5,1}+18 T_{5,2}+9 T_{5,3}+6 T_{5,4}+5 T_{5,5}\Bigr).
\label{mom_eq}
\end{align}

The above moments can be numerically computed by substituting  equation \ref{R5_P} into the integrand of equation \ref{mom_int} with $L = -ln (w)$.
The six-loop coefficients $T_{5,0}^{Pad\acute{e}}-T_{5,5}^{Pad\acute{e}}$  are determined by substituting these numerical values in equation  \ref{mom_eq}. The APAP predictions are compared to the RGE predictions for $n_f=5$ flavours obtained from equations \ref{RG_coeff_ms}\,-\ref{RG_coeff_MM}.  We estimate the relative error of the APAP predictions of the coefficients $T_{5,1}-T_{5,5}$ by computing $ \delta T_{5,i}^{ Pad\acute{e}} \equiv (T_{5,i}^{Pad\acute{e}} - T_{5,i}) / T_{5,i}$.  Our predictions for the $\overline{\text{MS}} $,  SI, OS, and the miniMOM schemes are given in tables \ref{tab2}-\ref{tab5}.
\begin{table}[H]
\renewcommand{\arraystretch}{1.5}
\begin{center}
\begin{tabular}{|c|c|c|c|c|c|c|}
\hline
 $\overline{\text{MS}} $& $T_{5,0}$ & $T_{5,1}$ & $T_{5,2}$ & $T_{5,3}$ & $T_{5,4}$ & $T_{5,5}$\\ \hline \hline
 $Pad\acute{e}$ & -110686 & 12383.7 & 100812 & 39138.2 & 4417.21 & 199.466  \\ \hline 
RGE & - & 15570.1 & 102046 & 38766.7 & 4613.1 & 155.197\\ \hline
$\delta T_{5,i}^{ Pad\acute{e}}$ & - & 20.5$\%$  &  1.2$\%$ & 1.0$\%$ & 4.2$\% $ & $28.5\%$ \\ \hline
\end{tabular}
\end{center}
\caption{The APAP predictions of the coefficients $R_5$ using the $S[1|3]$ Pad\'e approximant in the  $\overline{\text{MS}} $ scheme with errors. }
\label{tab2}
\end{table}
\begin{table}[H]
\renewcommand{\arraystretch}{1.5}
\begin{center}
\begin{tabular}{|c|c|c|c|c|c|c|}
\hline
 SI & $T_{5,0}$ & $T_{5,1}$ & $T_{5,2}$ & $T_{5,3}$ & $T_{5,4}$ & $T_{5,5}$\\ \hline \hline
 $Pad\acute{e}$ & -110378 & -4477.15 & 92857.5 & 37021 & 4684.3 &139.191  \\ \hline 
RGE & - & -3333.48 & 92400.3 & 37279.7 & 4530.19 & 155.197\\ \hline
$\delta T_{5,i}^{ Pad\acute{e}}$ & - & 34.3$\%$  &  0.5$\%$ & 0.7$\%$ & 3.4$\% $ & 10.3$\%$ \\ \hline
\end{tabular}
\end{center}
\caption{The APAP predictions of the coefficients $R_5$ using the $S[1|3]$ Pad\'e approximant in the  SI scheme with errors. }
\label{tab3}
\end{table}
\begin{table}[H]
\renewcommand{\arraystretch}{1.5}
\begin{center}
\begin{tabular}{|c|c|c|c|c|c|c|}
\hline
 OS & $T_{5,0}$ & $T_{5,1}$ & $T_{5,2}$ & $T_{5,3}$ & $T_{5,4}$ & $T_{5,5}$\\ \hline \hline
 $Pad\acute{e}$ & -105076 & 19853.1 & 104463 & 39648.4 & 4417.98 & 188.272  \\ \hline 
RGE & - & 20907.1 & 104628 & 39147.8 & 4613.1 & 155.197\\ \hline
$\delta T_{5,i}^{ Pad\acute{e}}$ & - & 5.0$\%$  &  0.2$\%$ & 1.3$\%$ & 4.2$\% $ & 21.3$\%$ \\ \hline
\end{tabular}
\end{center}
\caption{The APAP predictions of the coefficients $R_5$ using the $S[1|3]$ Pad\'e approximant in the OS scheme with errors. }
\label{tab4}
\end{table}
\begin{table}[H]
\renewcommand{\arraystretch}{1.5}
\begin{center}
\begin{tabular}{|c|c|c|c|c|c|c|}
\hline
 miniMOM & $T_{5,0}$ & $T_{5,1}$ & $T_{5,2}$ & $T_{5,3}$ & $T_{5,4}$ & $T_{5,5}$\\ \hline \hline
 $Pad\acute{e}$ & -72320.9 & -118501 & 10748.5 & 23659 & 2586.12 & 197.956 \\ \hline 
RGE & - & -102578 & 9429.12 & 25325.4 & 4111.7 & 155.197\\ \hline
$\delta T_{5,i}^{ Pad\acute{e}}$ & - & 15.5$\%$  &  14$\%$ & 6.6$\%$ & 37$\% $ & 27.6$\%$ \\ \hline
\end{tabular}
\end{center}
\caption{The APAP predictions of the coefficients $R_5$ using the $S[2|2]$ Pad\'e approximant in the  miniMOM scheme with errors. }
\label{tab5}
\end{table}

We observe that there is a good agreement between the APAP and RGE predictions of the $T_{5,1}-T_{5,5}$ coefficients in all the four schemes.  This gives us the required confidence in the APAP predictions of these coefficients,  and suggests the credibility of  the APAP prediction of the coefficient  $T_{5,0}$ which is inaccessible  through the RG-invariance.   

The predictions of the coefficients $R_i$ for  $i=5-9$ of the perturbative expansions in the  $\overline{\text{MS}} $,  SI,  OS, and miniMOM  schemes are given in table \ref{tab6},  where the coefficients for $i=6-9$ are  computed  using the $[3|2],[4|2],[5|2],[6|2]$ Pad\'e approximants.   We compare these predictions to that predicted by RGE in the third column.   The coefficient $T_{5,0}$ is not accessible through the RGE predictions.  Therefore,  for the prediction of the coefficient $R_5^{ RGE}$,  we assume the $T_{5,0}$ of the RGE identical to the $T_{5,0}$ of the Pad$\acute{e}$ prediction.
 \begin{table}[H]
\renewcommand{\arraystretch}{1.5}
\begin{center}
\begin{tabular}{|c|c|c|c|c|c|c|c|}
\hline
Scheme  & $R_5^{ Pad\acute{e}}$ &$R_5^{ RGE}$&$\delta R_5^{ Pad\acute{e}}$ & $R_6^{ Pad\acute{e}}$ & $R_7^{ Pad\acute{e}}$ & $R_8^{ Pad\acute{e}}$ & $R_9^{ Pad\acute{e}}$ \\ \hline \hline
$\overline{\text{MS}} $ & -86422.1  & -87826.6 & 1.6 $\%$&-285341 & $3.3315\times10^6$ & $4.83632\times10^7$ & $1.99309\times10^8$ \\ \hline 
SI &-86387.5 &-87191.2 & 0.9 $\%$&-266735 & $3.53838\times10^6$ & $4.82802\times10^7$ & $1.75429\times10^8$\\ \hline
OS &  -84249.7 &-84689.2 & 0.5 $\%$&-280858 & $3.14976\times10^6$ & $4.5776\times10^7$ & $1.87582\times10^8$ \\ \hline
miniMOM & 2761.63 & -8253.07 & 133$\%$ & 496045 & $2.82424\times10^6 $  & $-1.66589\times10^7$  &$-2.93284\times10^8$ \\ \hline
\end{tabular}
\end{center}
\caption{The APAP predicted  values of $R_{5-9}$  at $\mu=M_H$ and RGE predicted value of $R_5$ at $\mu=M_H$ with their relative error $\delta R_5^{ Pad\acute{e}}$, where $\delta R_5^{ Pad\acute{e}}=(R_5^{ RGE}-R_5^{ Pad\acute{e}})/R_5^{ RGE} \times 100$ in the $\overline{\text{MS}} $, SI, OS, and miniMOM schemes.}
\label{tab6}
\end{table}

We notice from table \ref{tab6} that the overall coefficient $R_5$ is negative in the $\overline{\text{MS}} $, SI, and OS schemes,  and is in an agreement with that predicted from the use of RGE.   This can be seen in the fourth column where the uncertainty $\delta R_5^{ Pad\acute{e}}$ between the  Pad$\acute{e}$ and RGE predictions is computed.    The largest uncertainty is 1.6 $\%$ in the $\overline{\text{MS}} $ scheme.   However,  the coefficient $R_5$ is positive in the miniMOM scheme,  and the uncertainty is quite large.  Therefore,  the miniMOM scheme is not working well in the APAP formalism, and its predictions are not reliable.     This particular feature of the miniMOM scheme continues to hold  in the PBA formalism as well.

The asymptotic errors $\delta_i$,  defined in equation \ref{error_for},  on our predictions presented  in \ref{tab6} are computed in the four different schemes,  and given in table \ref{tab7}. 
\begin{table}[H]
\renewcommand{\arraystretch}{1.5}
\begin{center}
\begin{tabular}{|c|c|c|c|c|c|}
\hline
Scheme  & $\delta_5$ & $\delta_6$ & $\delta_7$ & $\delta_8$ & $\delta_9$ \\ \hline \hline
$\overline{\text{MS}} $ &  $9.89\times10^{-19}$ & $-2.40\times10^{-12}$ &  $-2.40\times10^{-12}$  &  $-2.40\times10^{-12}$ & $-2.40\times10^{-12}$\\ \hline 
SI & $9.66\times10^{-19 }$ & $-2.66\times10^{-12}$ & $-2.37\times10^{-12}$ & $-2.37\times10^{-12}$ & $-2.37\times10^{-12}$\\ \hline
OS & $-9.89\times10^{-19}$ & $-2.40\times10^{-12}$ & $-2.40\times10^{-12}$ & $-2.40\times10^{-12}$ & $-2.40\times10^{-12}$\\ \hline
miniMOM & $-2.19\times10^{-11}$  & $-2.19\times10^{-11}$ & $-2.19\times10^{-11}$ & $-2.19\times10^{-11} $ & $-2.19\times10^{-11}$ \\ \hline
\end{tabular}
\end{center}
\caption{Asymptotic errors for $R_{5-9}$ in the $\overline{\text{MS}} $, SI, OS, and miniMOM schemes.}
\label{tab7}
\end{table}

The prediction of  the  coefficients $T_{4,i}$  and their  relative errors $\delta T_{4,i}^{ Pad\acute{e}}\equiv (T_{4,i}^{Pad\acute{e}} - T_{4,i}) / T_{4,i} \times 100$ are obtained using  the $[0|3]$ Pad\'e approximant.  These predictions  in the $\overline{\text{MS}} $, SI, OS, and miniMOM schemes are given in tables \ref{tab8}-\ref{tab11}.

\begin{table}[H]
\renewcommand{\arraystretch}{1.5}
\begin{center}
\begin{tabular}{|c|c|c|c|c|c|}
\hline
 $\overline{\text{MS}} $& $T_{4,0}$ & $T_{4,1}$ & $T_{4,2}$ & $T_{4,3}$ & $T_{4,4}$\\ \hline \hline
 $Pad\acute{e}$ & -1543.32 & 14291.6 & 9654.97 & 1577.93 & 67.477   \\ \hline 
Known &  -453.772 & 15627.2 & 9595.85 & 1574.97 & 67.4771\\ \hline
$\delta T_{4,i}^{ Pad\acute{e}}$ &  240.1$\%$  &  8.5$\%$ & 0.6$\%$ & 0.2$\% $ & 0.0002$\%$ \\ \hline
\end{tabular}
\end{center}
\caption{The APAP predictions of the coefficients $R_4$ in the $\overline{\text{MS}}$ scheme  with errors. }
\label{tab8}
\end{table}

\begin{table}[H]
\renewcommand{\arraystretch}{1.5}
\begin{center}
\begin{tabular}{|c|c|c|c|c|c|}
\hline
SI & $T_{4,0}$ & $T_{4,1}$ & $T_{4,2}$ & $T_{4,3}$ & $T_{4,4}$\\ \hline \hline
 $Pad\acute{e}$ & -2911.26 & 12890.6 & 9263.72 & 1549.09 & 67.4769  \\ \hline 
Known &  -1891.18 & 14042.3 & 9217.44 & 1546.13 & 67.4771\\ \hline
$\delta T_{4,i}^{ Pad\acute{e}}$ &  53.9$\%$  &  8.2$\%$ & 0.5$\%$ & 0.2$\% $ & 0.0003$\%$ \\ \hline
\end{tabular}
\end{center}
\caption{The APAP predictions of the coefficients $R_4$ in the SI scheme  with errors. }
\label{tab9}
\end{table}

\begin{table}[H]
\renewcommand{\arraystretch}{1.5}
\begin{center}
\begin{tabular}{|c|c|c|c|c|c|}
\hline
OS & $T_{4,0}$ & $T_{4,1}$ & $T_{4,2}$ & $T_{4,3}$ & $T_{4,4}$\\ \hline \hline
 $Pad\acute{e}$ & -922.469 & 14873.9 & 9742.49 & 1577.93 & 67.477  \\ \hline 
Known &  -5.68386 & 16064.3 & 9695.27 & 1574.97 & 67.4771\\ \hline
$\delta T_{4,i}^{ Pad\acute{e}}$ &  16129$\%$  &  7.4$\%$ & 0.5$\%$ & 0.2$\% $ & 0.0001$\%$ \\ \hline
\end{tabular}
\end{center}
\caption{The APAP predictions of the coefficients $R_4$ in the OS scheme  with errors. }
\label{tab10}
\end{table}

\begin{table}[H]
\renewcommand{\arraystretch}{1.5}
\begin{center}
\begin{tabular}{|c|c|c|c|c|c|}
\hline
miniMOM & $T_{4,0}$ & $T_{4,1}$ & $T_{4,2}$ & $T_{4,3}$ & $T_{4,4}$\\ \hline \hline
 $Pad\acute{e}$ & -8551.44 & -493.732 & 6798.1 & 1571.28 & 67.4771  \\ \hline 
Known &  -8939.29 & 401.918 & 6130.7 & 1400.56 & 67.4771\\ \hline
$\delta T_{4,i}^{ Pad\acute{e}}$ &  4.3$\%$  &  222.8$\%$ & 10.9$\%$ & 12.2$\% $ & 0.00004$\%$ \\ \hline
\end{tabular}
\end{center}
\caption{The APAP predictions of the coefficients $R_4$ in the miniMOM scheme  with errors. }
\label{tab11}
\end{table}

We note that the APAP formalism in this case is not able to reproduce the leading coefficient $T_{4,0}$ in  the $\overline{\text{MS}} $, SI,  and OS schemes.   In the case of the miniMOM scheme,  this happens with the coefficient  $T_{4,1}$.  However,  we also notice that the overall coefficient $R_4$ of the perturbative expansion  is in a good agreement with the APAP predictions as shown in table \ref{tab12}.  Moreover,  we shall show in the next section that the asymptotic Pad\'e-Borel  approximant  is capable of improving the APAP predictions of the coefficients  $T_{4,i}$,  therefore  reproducing  the known coefficient $R_4$ with a better accuracy.

\begin{table}[H]
\renewcommand{\arraystretch}{1.5}
\begin{center}
\begin{tabular}{|c|c|c|c|c|}
\hline
Scheme & $R_4$ & $R_4^{ Pad\acute{e}}$ & $\delta R_4$ & $\delta_4$ \\ \hline \hline
$\overline{\text{MS}} $   & -6957.06 & -7159.99 & 2.9 $\%$ & $9.88566\times 10^{-19}$ \\ \hline 
SI & -7019.12 & -7405.32 & 5.5$\%$ & $9.66238\times10^{-19}$ \\ \hline
OS & -6749.84 & -6878.79 & 1.9$\%$ & $9.88566\times10^{-19}$  \\ \hline
miniMOM & -7006.54 & -5807.67 & 17.1$\%$ & $2.72026\times10^{-17}$ \\ \hline
\end{tabular}
\end{center}
\caption{The APAP predicted values of $R_4$ at $\mu=M_H$ with relative error $\delta R_4$ and asymptotic error $\delta_4$,   where $\delta R_4=( R_4-R_4^{ Pad\acute{e}})/R_4 \times 100$.}
\label{tab12}
\end{table}

We observe that the accuracy obtained in the predictions of the  six-loop coefficients $T_{5,1}^{ Pad\acute{e}} -T_{5,5}^{ Pad\acute{e}}$  using the APAP method,  in particular  for the  $\overline{\text{MS}}$  and OS schemes  imply that  we can employ the APAP formalism with confidence elsewhere.  Therefore,  we  now use the APAP algorithm  to predict the coefficients  $\beta_5-\beta_9$ and $\gamma_5-\gamma_9$,  the higher-order loop  corrections to the $\beta$ and $\gamma$ functions.  These coefficients will be used to compute the  RGSPT function $S_5 (u)-S_9(u)$.

The $\beta$ function defined in equation  \ref{beta_6f} can be written as,

\begin{equation}
\beta(x)\,=\, - \beta_0 x^2 \sum_{i = 0}^\infty R_i x^i,
\end{equation}

where  $ R_i \equiv \beta_i/ \beta_0 $.   Using the already known values of  $\beta_0$ to  $\beta_4$ given in equation \ref{beta_nf},  we estimate the values of  $\beta_5,\beta_6,\beta_7,\beta_8$ and $\beta_9$  for $n_f=5$ using $[1|3],[3|2],[4|2],[5|2]$ and $[6|2]$ Pad\'e approximants respectively. The values of $\beta_5-\beta_9$ obtained are as follows, 

\begin{align}
 \beta_5=54.8149
\, ,\beta_6=61.6663
\, ,\beta_7=166.748
\, ,\beta_8=228.033
\, ,\beta_9=524.048.
\end{align}

The $\gamma_m$ function as defined in equation  \ref{beta_6f} can be written as

\begin{equation}
\gamma_m(x)=x\sum_{i = 0}^\infty R_i x^i,
\end{equation}
where $R_i=\gamma_i$. 

We use $[1|3],[3|2],[4|2],[5|2]$ and $[6|2]$  Pad\'e approximants  to predict the values of $\gamma_5,\gamma_6,\gamma_7,\gamma_8$ and $\gamma_9$ respectively using the already known values of $\gamma_0-\gamma_4$ for $n_f=5$. The predictions for $\gamma_5-\gamma_9$ are as follows,

\begin{align}
\gamma_5=659.411
\, ,\gamma_6=14271.8
\, ,\gamma_7=316090
\, ,\gamma_8=7.01049\times10^6
\, ,\gamma_9=1.55497\times10^8.
\end{align}

\begin{figure}[H]
  \centering
  % include first image
\subfigure[]{\includegraphics[width=0.45\textwidth]{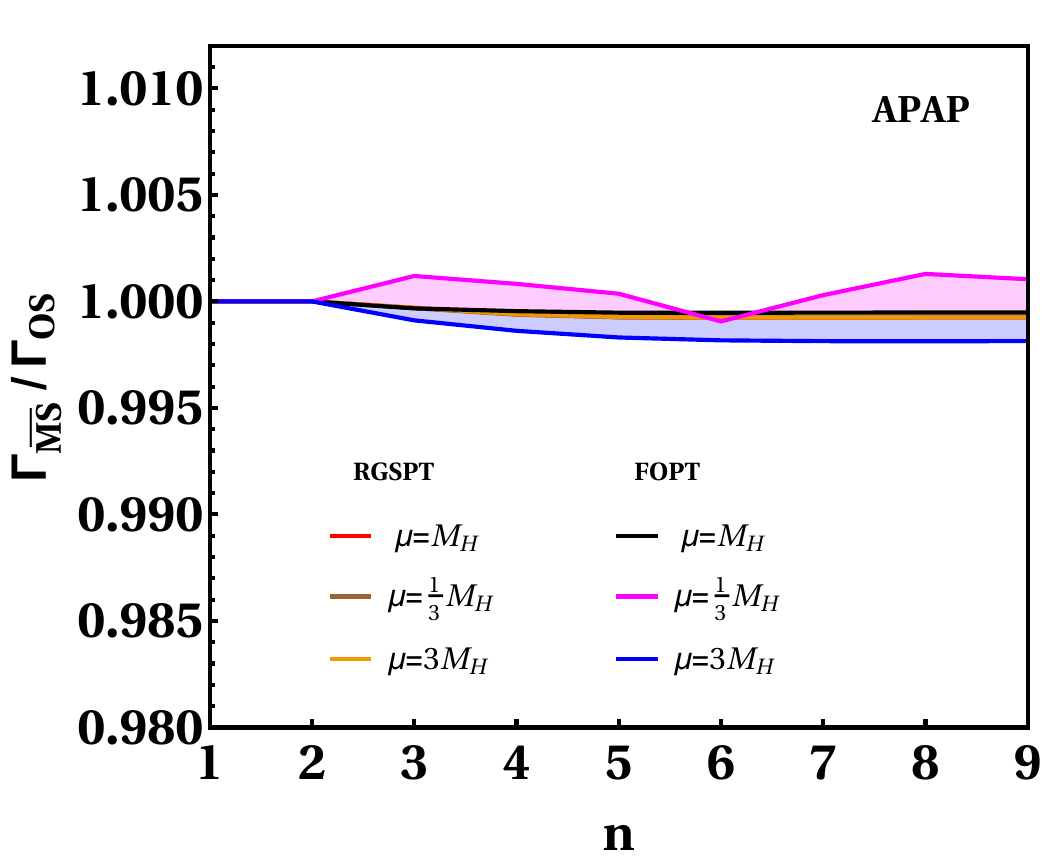}} 
\subfigure[]{\includegraphics[width=0.45\textwidth]{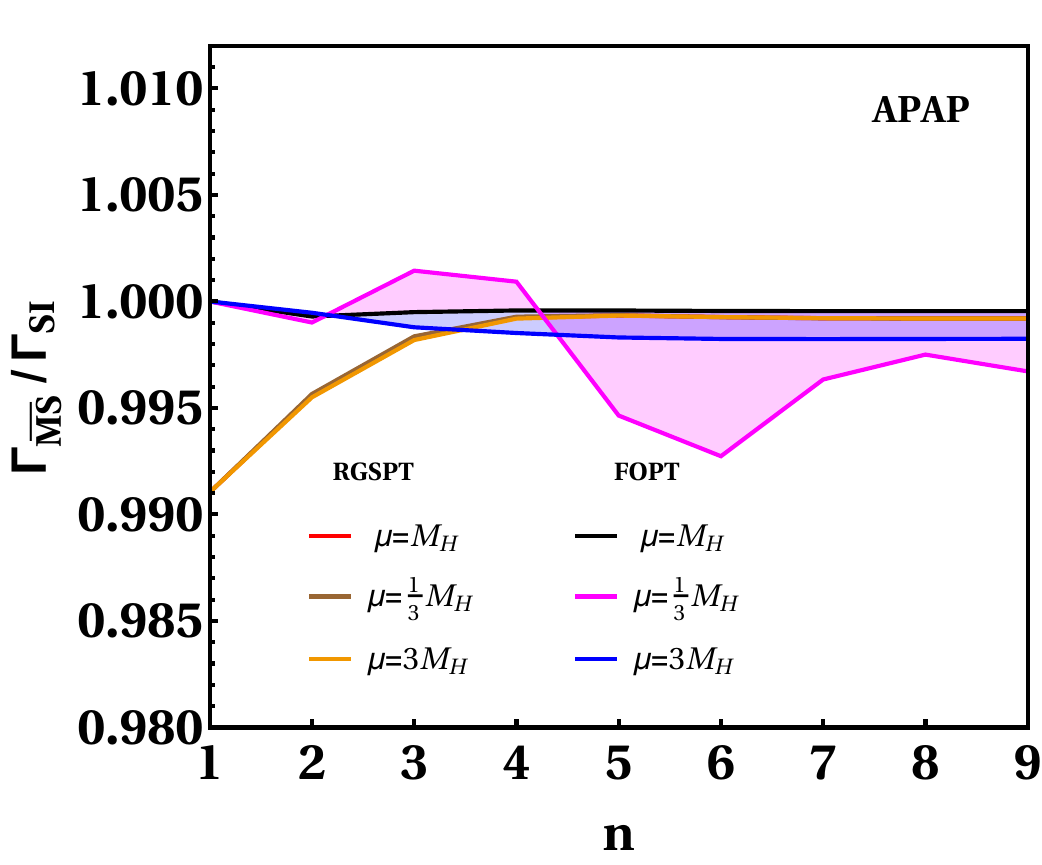}}   
\subfigure[]{\includegraphics[width=0.45\textwidth]{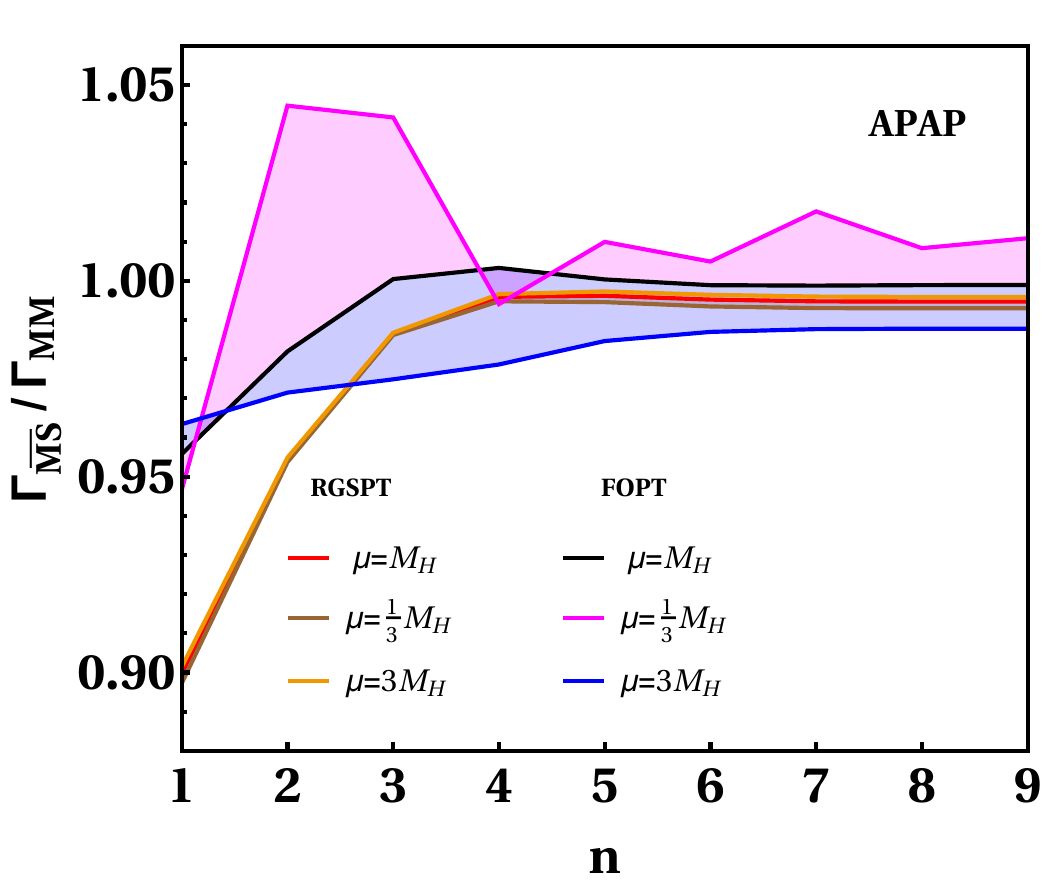}} 
   \caption{The variation of   $\Gamma_{\overline{\text{MS}}}/\Gamma_{\text{scheme}}$  at RG scales $\mu=\frac{1}{3}M_H, M_H$, and $3M_H$ in the  FOPT and RGSPT  in  the (a)SI (b)OS and (c)miniMOM schemes up to order $n=9$ .}
\label{pade_scale}
\end{figure}

Now we discuss the implications of the APAP determination of the coefficients $T_{n,k}$,  $\beta_n$ and $\gamma_n$ ($n= 5-9$) to the $\Gamma (H \rightarrow gg)$ decay width in the FOPT and RGSPT.     For a better understanding,  we  present the higher-order behaviour  of  the ratio of the Higgs to gluons decay width in two different RG schemes at different orders in the FOPT and RGSPT in figure \ref{pade_scale}.   The  $\overline{\text{MS}} $  and the OS schemes are showing very similar behaviour.  The RGSPT predictions reside within the FOPT predictions in this case.    In the case of the SI scheme,  this occurs beyond $n =3$.  The miniMOM scheme is again not performing well.  However,  it becomes stable within the framework of the RGSPT  beyond $n =3$.

We provide  order-by-order perturbative evaluation  of the $\Gamma (H \rightarrow gg)$ decay width up to $ \rm N^5LO $ at $\mu=M_H$ in the FOPT using the APAP formalism as,

\begin{align}
\frac{\Gamma_{ Pad\acute{e}}^{\overline{\text{MS}}}(M_H)}{\Gamma_{LO}(M_H)}\,=\,& 1+0.641657+0.196393 +0.0176989-0.0114448-\,\, \fbox{0.00742209},\\ \nonumber
\frac{\Gamma_{ Pad\acute{e}}^{SI}(M_H)}{\Gamma_{LO}(M_H)}\,=\,& 1+0.641657+0.197175 +0.0178975-0.0115469-\,\,\fbox{0.0074973},\\ \nonumber
\frac{\Gamma_{ Pad\acute{e}}^{OS}(M_H)}{\Gamma_{LO}(M_H)}\,=\,& 1+0.641657+0.196393 +0.0180594-0.0111039-\,\,\fbox{0.00716578},\\ \nonumber
\frac{\Gamma_{ Pad\acute{e}}^{MM}(M_H)}{\Gamma_{LO}(M_H)}\,=\,& 1+0.533291+0.0659834-0.0398094-0.0163109+\,\,\fbox{0.000251122}.
\end{align}

Similar predictions  of the $\Gamma (H \rightarrow gg)$ decay width up to $ \rm N^5LO $ at $\mu=M_H$ in RGSPT using the APAP formalism are,

\begin{align}
\frac{\Gamma_{ Pad\acute{e}}^{\overline{\text{MS}}}(M_H)}{\Gamma_{LO}(M_H)}\,=\,& 1+0.695577+0.263334 +0.0546671-0.00126852-\,\, \fbox{0.00527992},\\ \nonumber
\frac{\Gamma_{ Pad\acute{e}}^{SI}(M_H)}{\Gamma_{LO}(M_H)}\,=\,& 1+0.686965+0.253127 +0.0487537-0.00318015-\,\,\fbox{0.00542712},\\ \nonumber
\frac{\Gamma_{ Pad\acute{e}}^{OS}(M_H)}{\Gamma_{LO}(M_H)}\,=\,& 1+0.695577+0.263334 +0.055281-0.000655924-\,\,\fbox{0.00502101},\\ \nonumber
\frac{\Gamma_{ Pad\acute{e}}^{MM}(M_H)}{\Gamma_{LO}(M_H)}\,=\,& 1+0.884649+0.167738-0.0114242-0.0203406-\,\,\fbox{0.00587368}.
\end{align}

\section{Asymptotic Pad\'e-Borel  approximant improved  $H \rightarrow gg$ decay rate}
\label{B-APAP}
In this section,  we apply the APAP formalism to the Borel transform of the FOPT expansion of the Higgs to gluons decay width.   This method is found to have a better convergence,  and is referred to as   Pad\'e - Borel  approximant (PBA)  in literature \cite{Samuel:1995jc}.  This formalism is applied to the $H \rightarrow b \bar{b}$ decay rate in reference \cite{Boito:2021scm}.   In this work,  we apply the generalized Borel transform  to the  perturbative expansion given in equation  \ref{pert_exp}.

We first briefly review the formalism of the generalized Borel transform \cite{boas_rp}.  Let $\Psi (t)$ be a function given by,
\begin{equation}
\Psi (t)  = \sum_{n=0}^\infty \Psi_n t^n.
\end{equation}

For our purpose,  $\Psi (t)$ is a comparison function restricted by auxiliary conditions $ \Psi_n>0$  and     $  \lim_{n \to \infty}  \Psi_{n+1}/ \Psi_n =0 $.   We note that a comparison function $\Psi (t)$ is necessarily an entire function ensured by the ratio test of convergence.      A function $f$  is referred to as $\Psi$-type if there exists  the following relation,

\begin{equation}
|f (r \exp (i \theta))| \leq M \Psi (\tau r),
\label{psi_type}
\end{equation}

where $M$ and $\tau $ are some numbers.    The infimum of numbers $\tau$ for which equation \ref{psi_type} holds defines the class of functions  called $\Psi$-type $\tau$ \cite{boas_rp}.  The $\Psi$-type of a function  can be obtained from the coefficients in its power series expansion using the Nachbin's theorem \cite{boas_rp},
\begin{equation}
\text{ \bf Nachbin's theorem}:  \text{A function} ~~ f (z) = \sum_{n=0}^\infty f_n z^n  ~~ \text{is of}~~  \Psi  \text{-type} ~~ \tau ~~\text{if and only if} ~~   \lim_{n \to \infty}  |f_{n}/ \Psi_n|^{1/n} =\tau .
\end{equation}
 
Now we can define the generalized Borel transform given by,
\begin{equation}
B[f(z)](u) = \sum_n^\infty \frac{f_n}{\Psi_n} u^n.
\end{equation}

Since the function $f$ is  $\Psi$-type $\tau$,  the domain of convergence of the  function $B[f(z)] $ is $|u| \leq \tau$.  Moreover,  we can define \cite{boas_rp},
\begin{equation}
f(z) =  \frac{1}{2 \pi i} \oint\limits_{\Gamma}  \Psi (z u)\,  B[f(z)](u)  \,  \mathrm{d}u.
\end{equation}

The standard Borel transform is recovered by $\Psi (t) = e^t$.

We define the generalized Borel transform of the perturbative expansion  \ref{pert_exp}  as,
\begin{equation}
B[S] (u)= \sum_n^\infty \left(   \frac{d_1}{n!} + \frac{d_2}{n!^2} + \frac{d_3}{n!^3} + \frac{d_5}{n!^5}   \right)  R_n  u^n,
\label{gen_borel}
\end{equation}

where $d_{1,2,3, 5}$ are the scheme-dependent  real constants  given in  table \ref{tab_di}.  This is a phenomenological model in the sense that it reproduces the coefficient $R_{4}$ of the series  \ref{pert_exp} in particular,  when the APAP formalism is applied to this Borel transform.  The coefficient $R_3$ cannot be reproduced  through the APAP formalism since  there are three unknowns $A,a$ and $b$ to be fixed through  the two known coefficients of the $R_{1,2}$ of the series  \ref{pert_exp}.

 \begin{table}[H]
\begin{center}
   \begin{tabular}{|c|c|c|c|c|} \hline
Schemes & $d_1$      &      $d_2$               &     $d_3$   &   $d_5$    \\  \hline \hline
$\overline{\text{MS}} $ & $0.5$      &      $1.5$               &     $0$   &   $1.2$    \\   \hline
OS & $1$      &      $0$               &     $1.623$   &   $0$    \\  \hline
SI & $0.87$      &      $0$               &     $1.6$   &   $0$    \\  \hline
miniMOM & $0.68$      &      $1.5$               &     $0$   &   $1.2$    \\  \hline
\end{tabular} 
\end{center}
\caption{The numerical values of the constants $d_{1,2,3,5}$.}
   \label{tab_di}
   \end{table}
   
The APAP formalism is now applied  to the generalized Borel transformation defined in equation \ref{gen_borel}.   We use the $S[1,2]$   Pad\'e approximant to predict the coefficient $R_4$,  and $S[1,3]$ $S[2,2]$   for the coefficient $R_5$.   The unknown constants of the error given in equation \ref{error_for} are chosen to be  $b=-a$.   Moreover, our predictions are stable for the large values of the constants $a$ and $b$,  such as $10^6$.  This means that as the constants $a$ and $b$ approach a large value,  our predictions become independent of these large values.  This also means from equation \ref{error_for} that the error on our predictions practically vanishes since the constant $A$ is independent of the values of  $a$ and $b$  for the choice $b=-a$.   Thus,  from equation  \ref{error_for}  for very large values of  $a$ and $b$  we have,

\be 
R_{N+M+1}^{Pad\acute{e}} = R_{N+M+1}.
\ee

The PBA predictions of the coefficient $R_4$ in the  $\overline{\text{MS}} $, SI, OS, and miniMOM schemes for  $n_f=5$ flavours  are given in table \ref{pba_ms_r4}.  The very first observation is the improvement of the  prediction of the coefficient $T_{4,0}$ over that of the APAP predictions.  This improvement continues even in the SI and OS schemes as shown in tables \ref{pba_si_r4} and \ref{pba_os_r4}. 

\begin{table}[H]
\renewcommand{\arraystretch}{1.5}
\begin{center}
\begin{tabular}{|c|c|c|c|c|c|}
\hline
 $\overline{\text{MS}} $& $T_{4,0}$ & $T_{4,1}$ & $T_{4,2}$ & $T_{4,3}$ & $T_{4,4}$\\ \hline \hline
 PBA & -559.149 & 15801.3 & 9837.78 & 1579.97 & 68.6471    \\ \hline 
Known &  -453.772 & 15627.2 & 9595.85 & 1574.97 & 67.4771\\ \hline
$\delta T_{4,i}$ &  23.2$\%$  &  1.1$\%$ & 2.5$\%$ & 0.3$\% $ & 1.7$\%$ \\ \hline
\end{tabular}
\end{center}
\caption{The PBA predictions of the coefficients $R_4$ in the  $\overline{\text{MS}} $ scheme with errors where, $ \delta T_{4,i}^{PBA} \equiv (T_{4,i}^{PBA} - T_{4,i}) / T_{4,i} \times 100$ and $i=0-4$. }
\label{pba_ms_r4}
\end{table}

\begin{table}[H]
\renewcommand{\arraystretch}{1.5}
\begin{center}
\begin{tabular}{|c|c|c|c|c|c|}
\hline
SI & $T_{4,0}$ & $T_{4,1}$ & $T_{4,2}$ & $T_{4,3}$ & $T_{4,4}$\\ \hline \hline
 PBA & -1482.92 & 14633.4 & 9574.99 & 1562.99 & 69.4497  \\ \hline 
Known &  -1891.18 & 14042.3 & 9217.44 & 1546.13 & 67.4771\\ \hline
$\delta T_{4,i}^{PBA}$ &  21.6$\%$  &  4.2$\%$ & 3.9$\%$ & 1.1$\% $ & 2.9$\%$ \\ \hline
\end{tabular}
\end{center}
\caption{The PBA predictions of the coefficients $R_4$ in the  SI scheme  with errors. }
\label{pba_si_r4}
\end{table}

\begin{table}[H]
\renewcommand{\arraystretch}{1.5}
\begin{center}
\begin{tabular}{|c|c|c|c|c|c|}
\hline
OS & $T_{4,0}$ & $T_{4,1}$ & $T_{4,2}$ & $T_{4,3}$ & $T_{4,4}$\\ \hline \hline
 PBA & -5.1548 & 16851.1 & 10232.9 & 1626.21 & 70.7353  \\ \hline 
Known &  -5.68386 & 16064.3 & 9695.27 & 1574.97 & 67.4771\\ \hline
$\delta T_{4,i}^{PBA}$ &  9.3$\%$  &  4.9$\%$ & 5.5$\%$ & 3.3$\% $ & 4.8$\%$ \\ \hline
\end{tabular}
\end{center}
\caption{The PBA predictions of the coefficients $R_4$ in the OS scheme  with errors. }
\label{pba_os_r4}
\end{table}

In the case of the miniMOM scheme as well,  we notice a good improvement in the prediction of the coefficient $T_{4,1}$ over the APAP prediction of the same coefficient as shown in table \ref{pba_mom_r4}.   Thus,  we can conclude that the PBA formalism is performing relatively  well for our perturbative expansion.

\begin{table}[H]
\renewcommand{\arraystretch}{1.5}
\begin{center}
\begin{tabular}{|c|c|c|c|c|c|}
\hline
miniMOM & $T_{4,0}$ & $T_{4,1}$ & $T_{4,2}$ & $T_{4,3}$ & $T_{4,4}$\\ \hline \hline
PBA & -8517.61 & 415.194 & 7204.86 & 1571.76 & 74.0561  \\ \hline 
Known &  -8939.29 & 401.918 & 6130.7 & 1400.56 & 67.4771\\ \hline
$\delta T_{4,i}^{PBA}$ &  4.7$\%$  &  3.3$\%$ & 17.5$\%$ & 12.2$\% $ & 9.8$\%$ \\ \hline
\end{tabular}
\end{center}
\caption{The PBA predictions of the coefficients $R_4$ in the miniMOM scheme  with errors. }
\label{pba_mom_r4}
\end{table}

In table \ref{pba_R4} we show our overall predictions of the coefficient $R_4$ in the four different schemes.  We observe a good agreement with that of already computed values of the coefficient $R_4$ in the four different schemes,  and conclude that the coefficient $R_4$  is better predicted by the PBA formalism in the $\overline{\text{MS}} $,  SI and OS schemes.

\begin{table}[H]
\renewcommand{\arraystretch}{1.5}
\begin{center}
\begin{tabular}{|c|c|c|c|c|}
\hline
Scheme & $R_4$ & $R_4^{PBA}$ & $\delta R_4$ & $\delta_4^{PBA}$ \\ \hline \hline
$\overline{\text{MS}} $   & -6957.06 & -6811.95 & 2.0 $\%$ & $-1.66573\times10^{-10}$ \\ \hline 
SI & -7019.12 & -6701.95 & 4.5$\%$ & $-1.01454\times10^{-18}$ \\ \hline
OS & -6749.84 & -6775.61 & 0.4$\%$ & $-9.31203\times10^{-13}$  \\ \hline
miniMOM & -7006.54 & -5929.47
 & 15.4$\%$ & $-6.74176\times10^{-12}$ \\ \hline
\end{tabular}
\end{center}
\caption{PBA predicted values of $R_4$ at $\mu=M_H$ with relative error and asymptotic error in the $\overline{\text{MS}} $, SI, OS, and miniMOM schemes. }
\label{pba_R4}
\end{table}

In a similar manner,   we  predict the coefficient $T_{5,i}$  for  $n_f=5$ flavours and compare it to that predicted by the RGE.  Our predictions for the $\overline{\text{MS}} $, SI, OS, and miniMOM schemes in tables \ref{pba_ms_r5}-\ref{pba_mom_r5}.

\begin{table}[H]
\renewcommand{\arraystretch}{1.5}
\begin{center}
\begin{tabular}{|c|c|c|c|c|c|c|}
\hline
 $\overline{\text{MS}} $& $T_{5,0}$ & $T_{5,1}$ & $T_{5,2}$ & $T_{5,3}$ & $T_{5,4}$ & $T_{5,5}$\\ \hline \hline
 PBA & -126999 & 11998.4 & 105921 & 40231.5 &4813.81 & 159.693 \\ \hline 
RGE & - & 15570.1 & 102046 & 38766.7 & 4613.1 & 155.197\\ \hline
$\delta T_{5,i}^{PBA}$ & - & 22.9$\%$  & 3.8$\%$ & 3.8$\%$ & 4.4$\% $ & 2.9$\%$ \\ \hline
\end{tabular}
\end{center}
\caption{The PBA predictions of the coefficients $R_5$ using the $S[1|3]$ Pad\'e approximant in the  $\overline{\text{MS}} $ scheme with errors. }
\label{pba_ms_r5}
\end{table}

\begin{table}[H]
\renewcommand{\arraystretch}{1.5}
\begin{center}
\begin{tabular}{|c|c|c|c|c|c|c|}
\hline
 SI & $T_{5,0}$ & $T_{5,1}$ & $T_{5,2}$ & $T_{5,3}$ & $T_{5,4}$ & $T_{5,5}$\\ \hline \hline
 PBA & -150441 & -5962.74 & 113068 & 45018.4  & 5508.34 & 185.327  \\ \hline 
RGE & - & -3333.48 & 92400.3 & 37279.7 & 4530.19 & 155.197\\ \hline
$\delta T_{5,i}^{PBA}$ & - & 78.9$\%$  & 22.4$\%$ & 20.8$\%$ & 21.6$\% $ & 19.4$\%$ \\ \hline
\end{tabular}
\end{center}
\caption{The PBA predictions of the coefficients $R_5$ using the $S[2|2]$ Pad\'e approximant in the  SI scheme with errors. }
\label{pba_si_r5}
\end{table}

\begin{table}[H]
\renewcommand{\arraystretch}{1.5}
\begin{center}
\begin{tabular}{|c|c|c|c|c|c|c|}
\hline
 OS & $T_{5,0}$ & $T_{5,1}$ & $T_{5,2}$ & $T_{5,3}$ & $T_{5,4}$ & $T_{5,5}$\\ \hline \hline
 PBA & -149714 & 12139.2 & 122438 & 45832.4 & 5490.66 & 180.862 \\ \hline 
RGE & - & 20907.1 & 104628 & 39147.8 & 4613.1 & 155.197\\ \hline
$\delta T_{5,i}^{PBA}$ & - & 41.9$\%$  &  17.0$\%$ & 17.1$\%$ & 19.0$\% $ & 16.5$\%$ \\ \hline
\end{tabular}
\end{center}
\caption{The PBA predictions of the coefficients $R_5$ using the $S[2|2]$ Pad\'e approximant in the OS scheme  with errors. }
\label{pba_os_r5}
\end{table}

\begin{table}[H]
\renewcommand{\arraystretch}{1.5}
\begin{center}
\begin{tabular}{|c|c|c|c|c|c|c|}
\hline
 miniMOM & $T_{5,0}$ & $T_{5,1}$ & $T_{5,2}$ & $T_{5,3}$ & $T_{5,4}$ & $T_{5,5}$\\ \hline \hline
 PBA & -69998 & -129042 & 12610.1 & 24334 & 2963.44 & 208.129 \\ \hline 
RGE & - & -102578 & 9429.12 & 25325.4 & 4111.7 & 155.197\\ \hline
$\delta T_{5,i}^{PBA}$ & - & 25.8$\%$  &  33.7$\%$ & 3.9$\%$ & 27.9$\% $ & 34.1$\%$ \\ \hline
\end{tabular}
\end{center}
\caption{The PBA predictions of the coefficients $R_5$ using the $S[2|2]$ Pad\'e approximant in the miniMOM scheme  with errors. }
\label{pba_mom_r5}
\end{table}

\begin{table}[H]
\renewcommand{\arraystretch}{1.5}
\begin{center}
\begin{tabular}{|c|c|c|c|c|c|c|c|}
\hline
Scheme  & $R_5^{PBA}$ & $R_5^{RGE}$& $\delta R_5^{PBA}$ & $R_6^{PBA}$ & $R_7^{PBA}$ & $R_8^{PBA}$ & $R_9^{PBA}$ \\ \hline \hline
$\overline{\text{MS}} $ & -100576 & -104140 & 3.4$\%$ & -579764 & $3.14089\times10^6$ & $1.28987\times10^8$ & $1.55196\times10^9$ \\ \hline 
SI & -120962 & -127255 & 4.9$\%$ & -927464 & $4.86118\times10^6$ & $3.10397\times10^8$ & $5.48341\times10^9$ \\ \hline
OS &  -117885 & -129328 & 8.8 $\%$& -893860 & $5.32098\times10^6$ & $3.12672\times10^8$ & $5.64975\times10^9$\\ \hline
miniMOM & 12553.1 & -5930.1 & 311.7$\%$ & 785972 & $4.58537\times10^6$ & $-9.36415\times10^7$ & $-1.96327\times10^9$\\ \hline
\end{tabular}
\end{center}
\caption{The PBA predicted values of $R_{5-9}$  at $\mu=M_H$ and RGE predicted value of $R_5$ at $\mu=M_H$ with their relative error $\delta R_5^{ PBA}$, where $\delta R_5^{ PBA}=(R_5^{ RGE}-R_5^{ PBA})/R_5^{ RGE} \times 100$ in the $\overline{\text{MS}} $, SI, OS, and miniMOM schemes.}
\label{pba_R5}
\end{table}

\begin{table}[H]
\renewcommand{\arraystretch}{1.5}
\begin{center}
\begin{tabular}{|c|c|c|c|c|c|}
\hline
Scheme  & $\delta_5^{PBA}$ & $\delta_6^{PBA}$ & $\delta_7^{PBA}$ & $\delta_8^{PBA}$ & $\delta_9^{PBA}$ \\ \hline \hline
$\overline{\text{MS}} $ & $5.70\times10^{-13}$ & $-1.67\times10^{-8}$ & $-1.67\times10^{-8}$ & $-1.67\times10^{-8}$ & $-1.67\times10^{-8}$\\ \hline 
SI & $-1.01\times10^{-10}$ & $-1.01\times10^{-18}$ & $-1.01\times10^{-18}$ & $-1.01\times10^{-10}$ & $-1.01\times10^{-18}$\\ \hline
OS & $-9.31\times10^{-19}$ & $-9.31\times10^{-19}$ & $-9.31\times10^{-19}$ & $-9.31\times10^{-19}$ & $-9.31\times10^{-19}$\\ \hline
miniMOM & $-6.74\times10^{-18}$ & $-6.74\times10^{-18}$ & $-6.74\times10^{-18}$ & $-6.74\times10^{-18}$ &  $-6.74\times10^{-18}$\\ \hline
\end{tabular}
\end{center}
\caption{The asymptotic errors for $R_{5-9}$ in the $\overline{\text{MS}} $, SI, OS, and miniMOM schemes.}
\label{pba_delta}
\end{table} 

\begin{table}[H]
\renewcommand{\arraystretch}{1.5}
\begin{center}
\begin{tabular}{|c|c|c|c|c|c|c|}
\hline
Scheme  & $\Delta T_{5,0}$ & $\Delta T_{5,1}$ &$\Delta T_{5,2}$ & $\Delta T_{5,3}$ & $\Delta T_{5,4}$ & $\Delta T_{5,5}$ \\ \hline \hline
$\overline{\text{MS}} $ & 16313.5 & 385.346 &  -5108.44  &  -1093.3 & -396.599 & 39.7737\\ \hline 
SI & 40063.5 & 1485.59 & -20210.4 & -7997.47 & -824.036 & -46.1366\\ \hline
OS & 44638.8 & 7713.91 & -17974.7 & -6184.05 & -1072.69 & 7.41051\\ \hline
miniMOM & -2322.97  & 10541.1 & -1861.61 & -675.004 & -377.319 & -10.1734 \\ \hline
\end{tabular}
\end{center}
\caption{Difference between the APAP predictions and PBA predictions for the coefficients $R_5$  at $\mu=M_H$ in the $\overline{\text{MS}} $, SI, OS, and miniMOM schemes where $\Delta T_{5,i}= T_{5,i}^{APAP}-T_{5,i}^{PBA}$.}
\label{tab25}
\end{table}

The predictions of the overall coefficients $R_i^{PBA}$ with $i=5-9$ in four different schemes are given in  table \ref{pba_R5}.  The predictions  for the coefficients $R_{6-9}$ are obtained  using the $S[3,2],S[4,2],S[5,2],S[6,2]$ Pad\'e approximants,  respectively in    the $\overline{\text{MS}} $,  SI,  OS, and miniMOM schemes.   The  asymptotic errors on the predictions of the overall coefficients $R_{5-9}$ in the $\overline{\text{MS}} $, SI, OS, and miniMOM schemes are extremely small, and are given in table \ref{pba_delta}.  The RGE prediction of the coefficient $R_{5}$  is obtained assuming that the coefficient $T_{5,0}$ in the RGE is identical to that of the PBA.

For a conservative estimate of the coefficient  $R_5$,  we associate the difference of the APAP and the PBA predictions of the coefficients $T_{5,i}$ as an uncertainty to the coefficient $R_5$.  This uncertainty is calculated by adding the errors on the coefficients $T_{5,i}$  given in table \ref{tab25} in quadrature.  This  results in the following values of the coefficient  $R_5$ at $\mu = \rm M_H$,
\begin{align}
R_5^{\overline{\text{MS}}}=& -86422.1 \pm 16456.8, \\ \nonumber 
R_5^{SI}=& -86387.5 \pm 40519.1, \\ \nonumber
R_5^{OS}=& -84249.7 \pm 45568.8, \\\nonumber
R_5^{MM}=& 2761.63 \pm 7238.89. \\ \nonumber
\end{align}

\begin{figure}[H]
  \centering
  % include first image
  \subfigure[]{\includegraphics[width=0.45\textwidth]{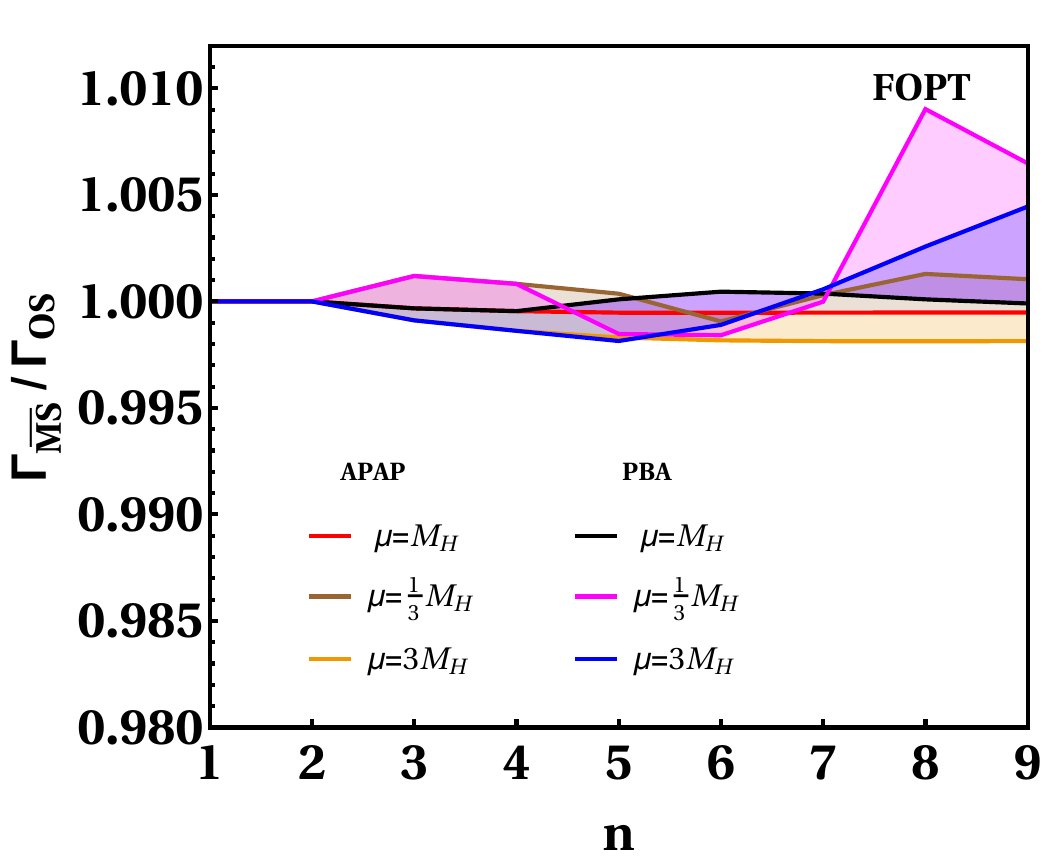}} 
    \subfigure[]{\includegraphics[width=0.45\textwidth]{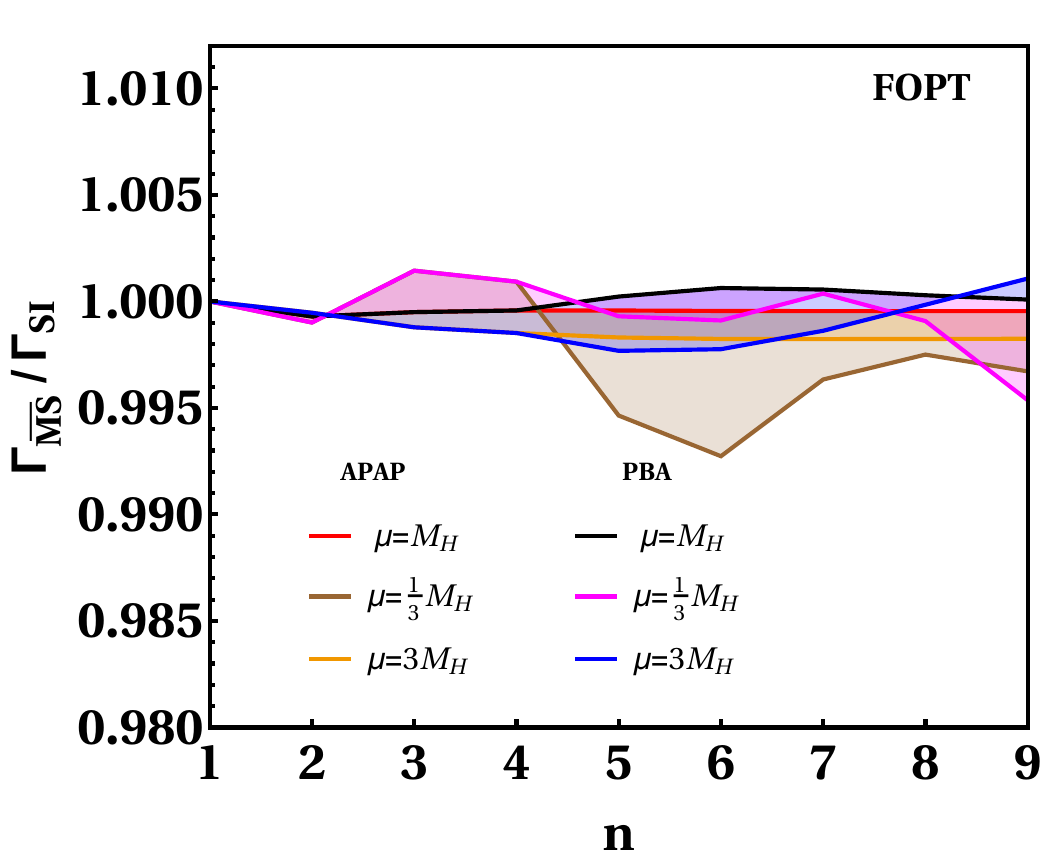}}   
     \subfigure[]{\includegraphics[width=0.45\textwidth]{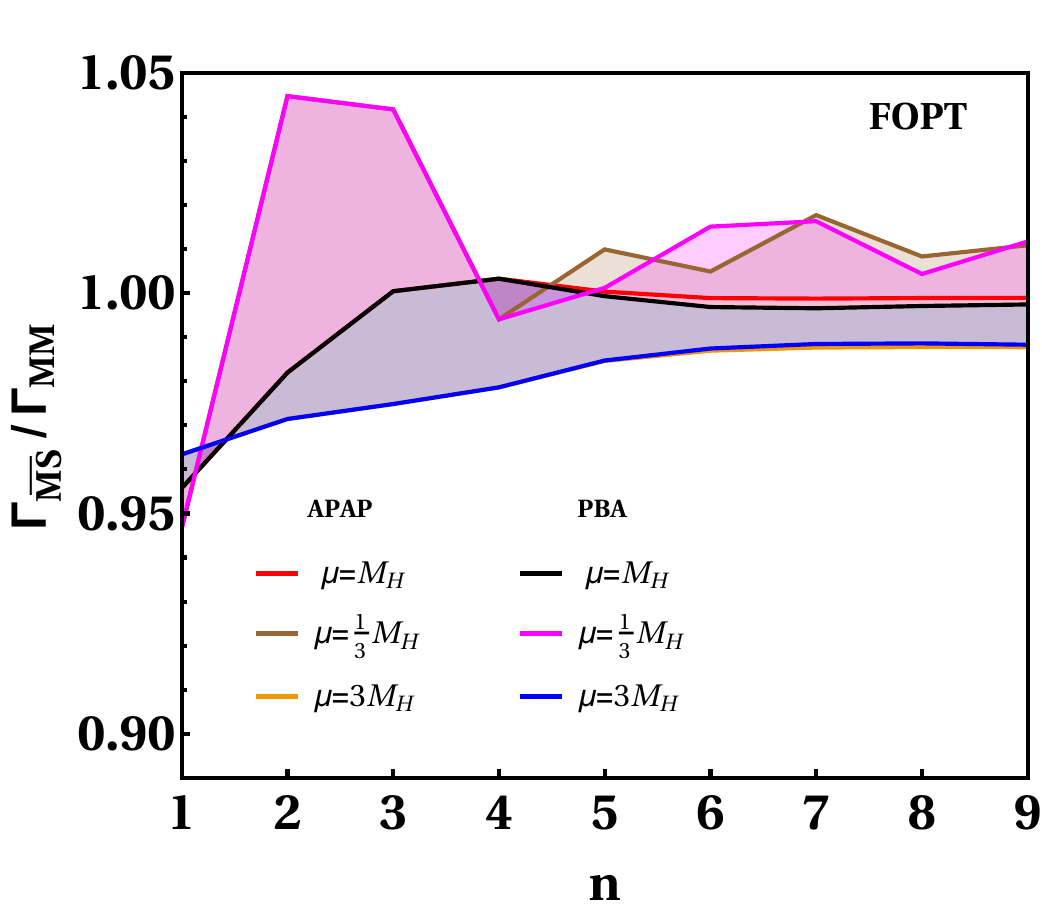}} 
    \caption{The variation of  $\Gamma_{\overline{\text{MS}}}/\Gamma_{\text{scheme}}$  at RG scales $\mu=\frac{1}{3}M_H, M_H$, and $3M_H$ in the  FOPT in  the (a)SI (b)OS and (c)miniMOM schemes up to order $n=9$ .}
\label{pba_scale_fopt}
\end{figure}

\begin{figure}[H]
  \centering
  % include first image
\subfigure[]{\includegraphics[width=0.45\textwidth]{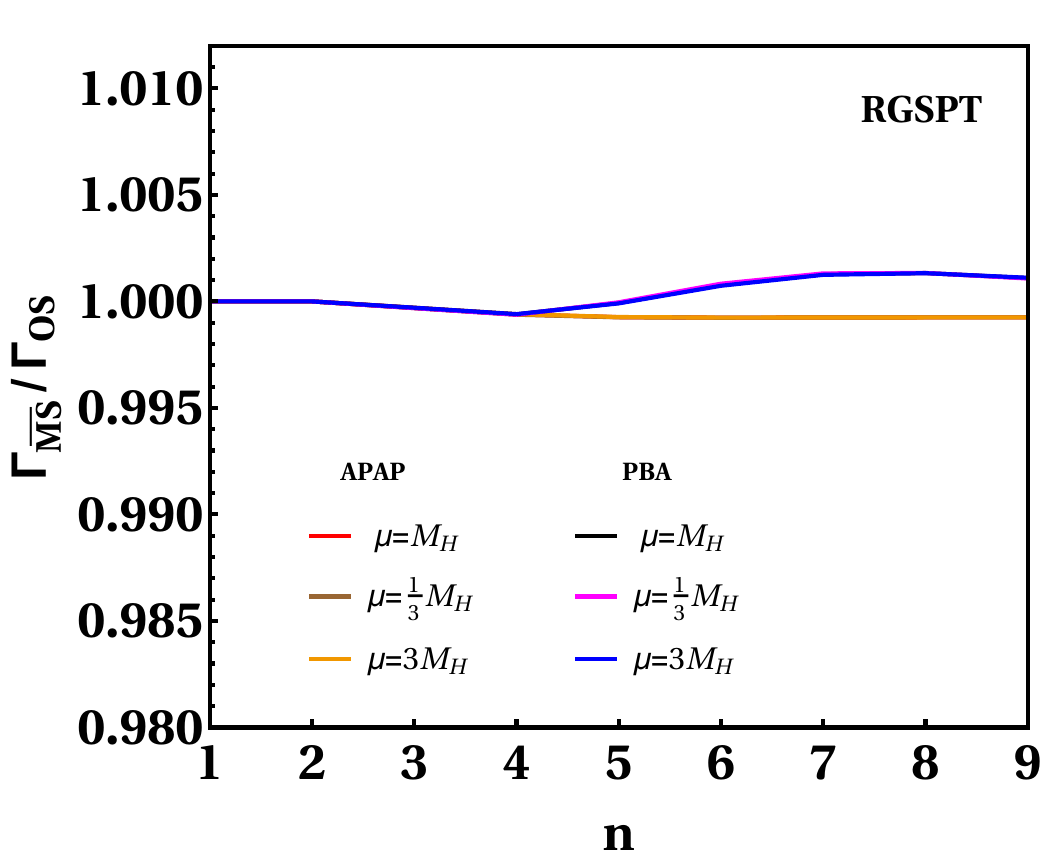}} 
\subfigure[]{\includegraphics[width=0.45\textwidth]{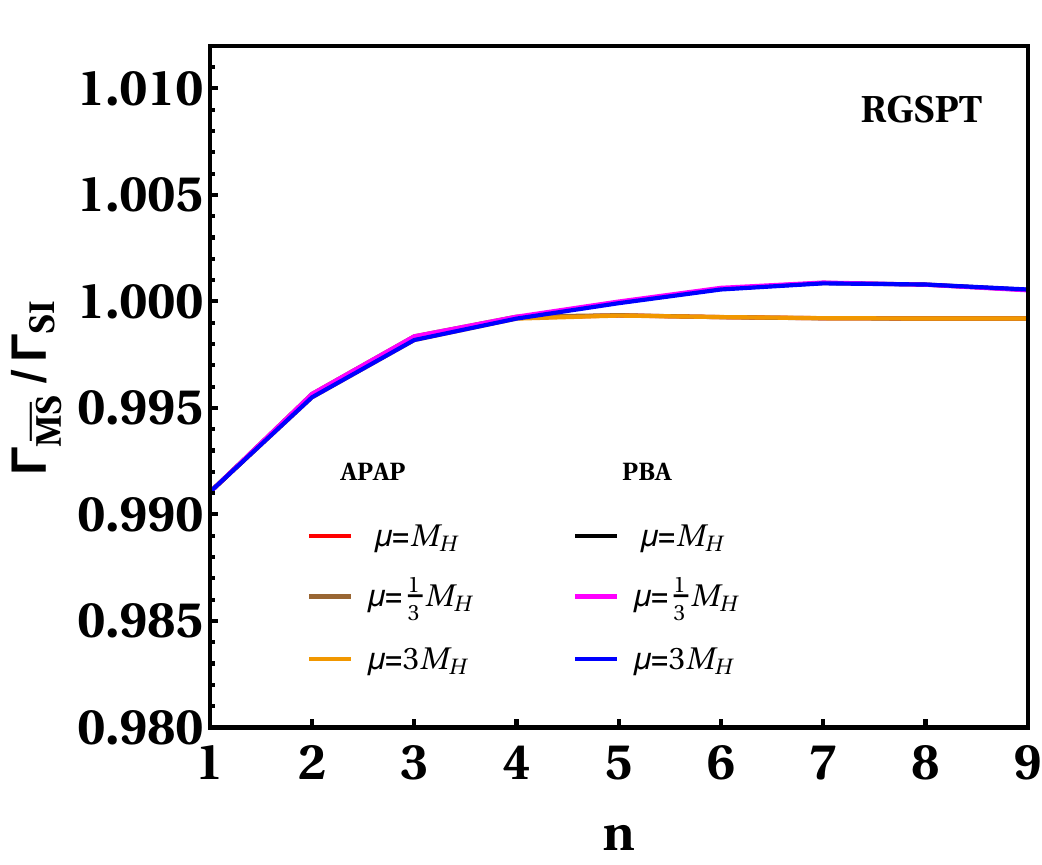}}   
\subfigure[]{\includegraphics[width=0.45\textwidth]{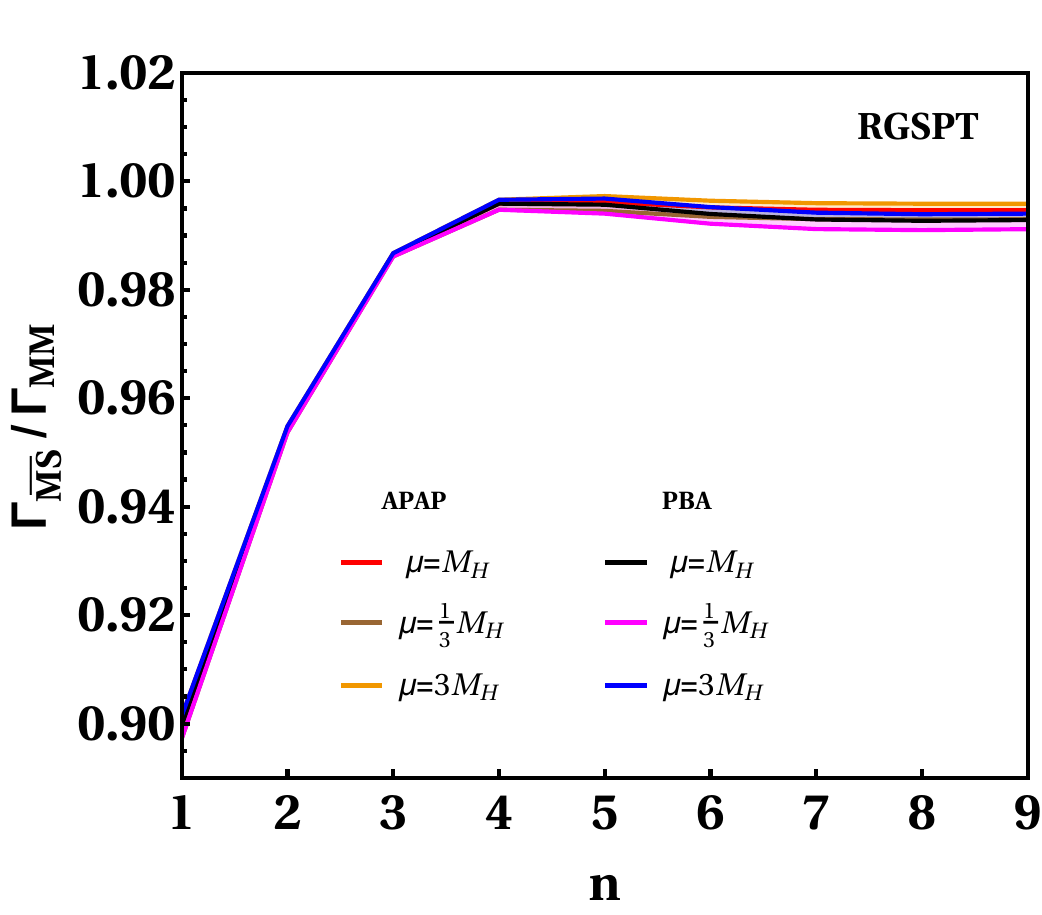}} 
   
\caption{The variation of   $\Gamma_{\overline{\text{MS}}}/\Gamma_{\text{scheme}}$  at RG scales $\mu=\frac{1}{3}M_H, M_H$, and $3M_H$ in the  RGSPT in  the (a)SI (b)OS and (c)miniMOM schemes up to order $n=9$ .}
\label{pba_scale_rgspt}
\end{figure}

The  higher-order behaviour of  the ratio of the Higgs to gluons decay width in two different RG schemes at different orders  predicted by the PBA formalism at  three different RG scales  in the FOPT is shown in figure \ref{pba_scale_fopt} along with that of the APAP formalism.  There is a good agreement between the predictions of the APAP and the PBA  formalisms for  the $\overline{\text{MS}} $,  SI,  and the OS schemes.    We show the  higher-order behaviour of  the ratio of the Higgs to gluons decay width in two different RG schemes at different orders   predicted by the PBA and the APAP formalisms  in the RGSPT in the $\overline{\text{MS}} $, SI, OS, and miniMOM schemes in figure \ref{pba_scale_rgspt} at three different scales.   As expected,  the RGSPT expansions are less sensitive to the RG scale dependence.

\begin{figure}[H]
  \centering
  % include first image
\subfigure[]{\includegraphics[width=0.45\textwidth]{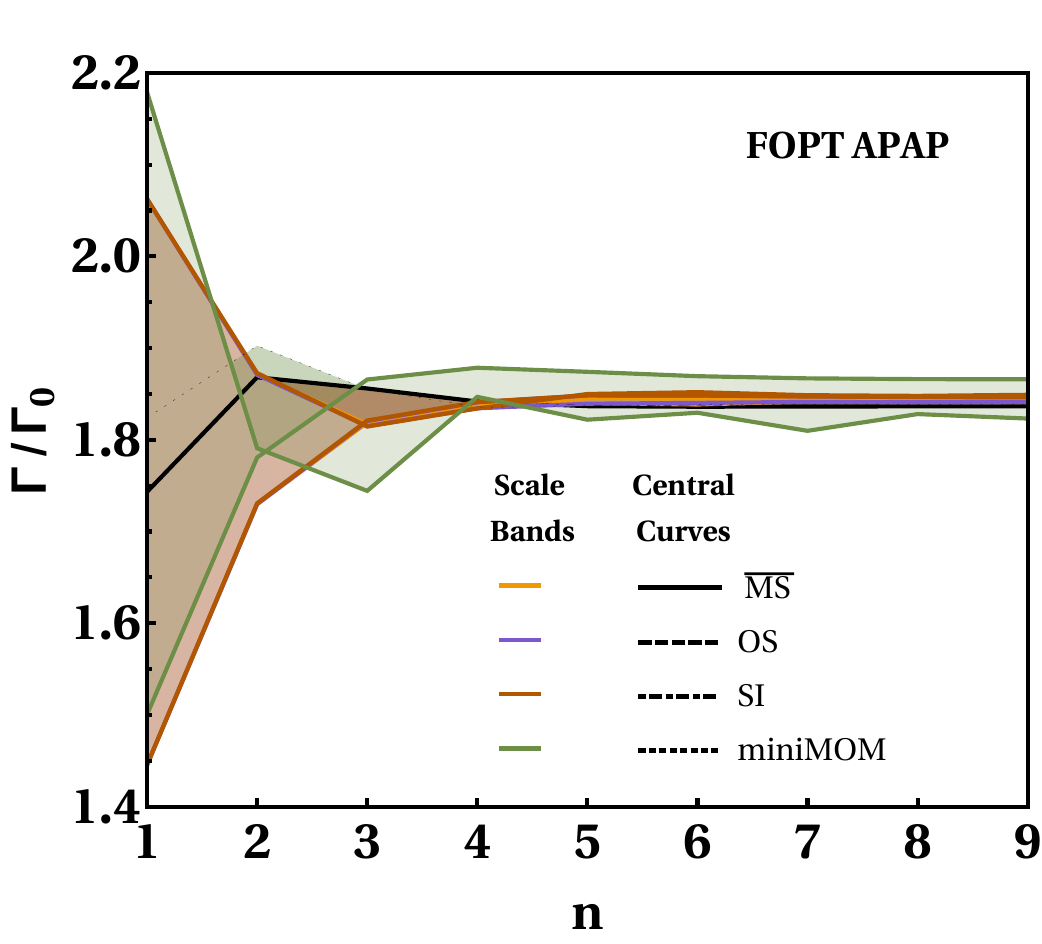}}
\subfigure[]{\includegraphics[width=0.46\textwidth]{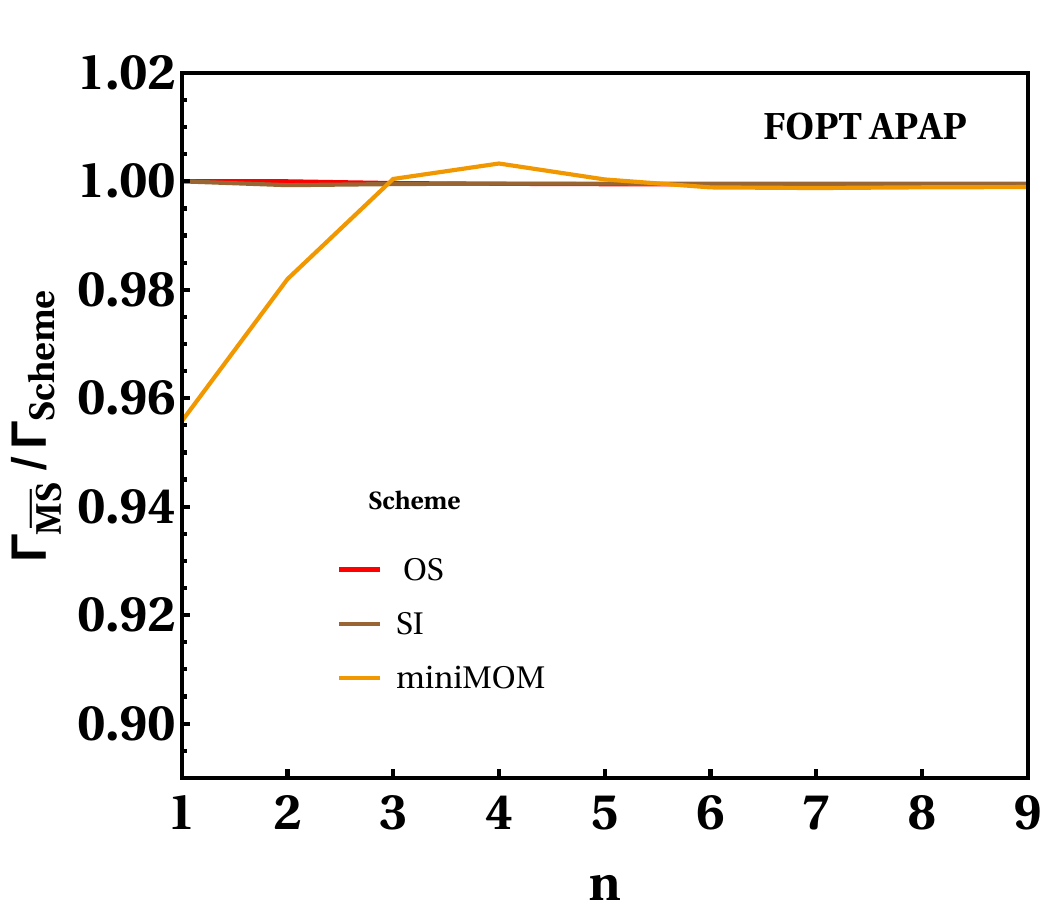}} 
   \subfigure[]{\includegraphics[width=0.45\textwidth]{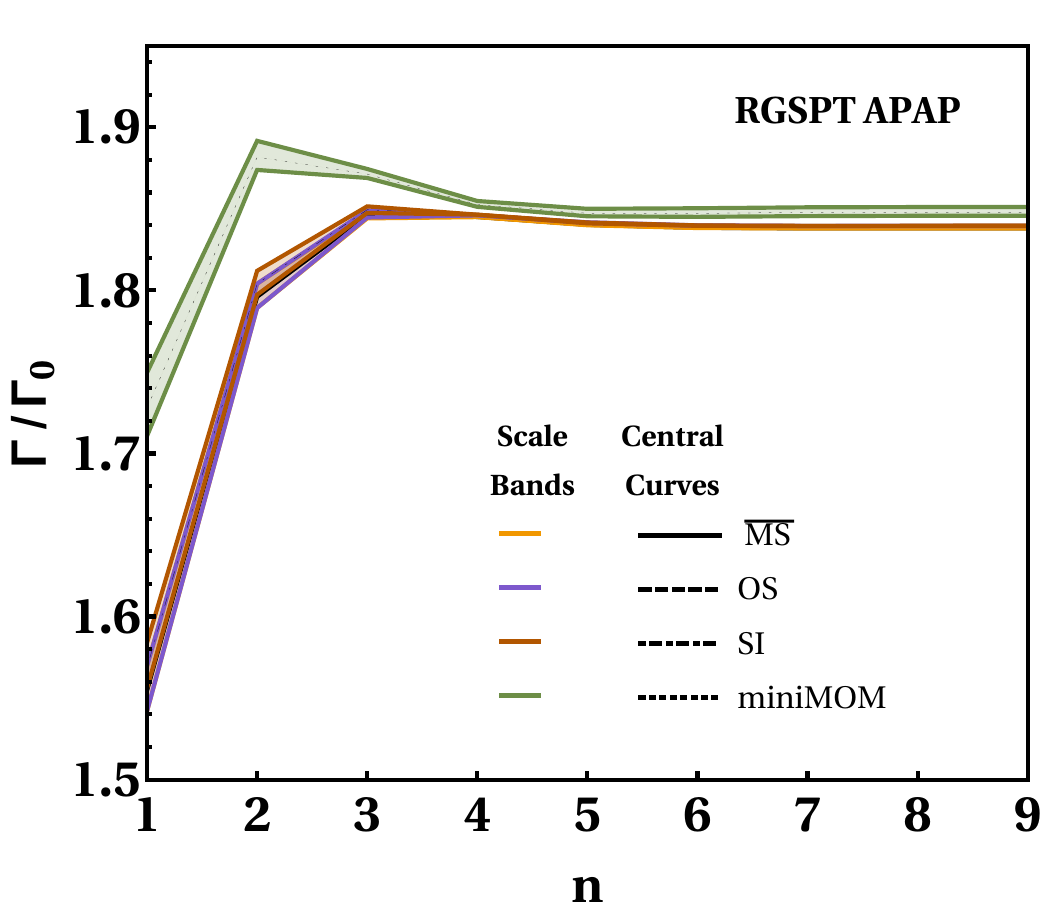}} 
 \subfigure[]{\includegraphics[width=0.46\textwidth]{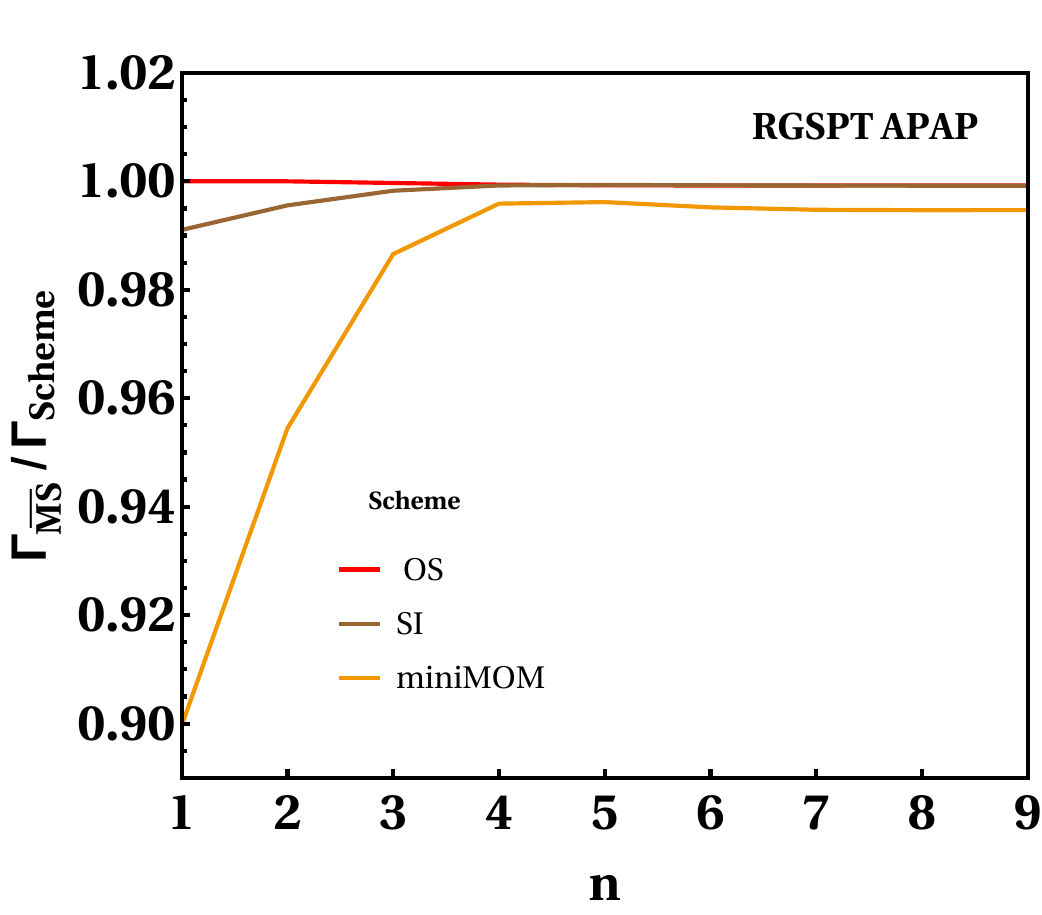}} 
\caption{The variation of   the normalised higgs to gluons decay width at RG scale $\mu= M_H$ with scale bands at $\mu=\frac{1}{3} M_H$ and $3 M_H$  in the (a)FOPT and the (b)RGSPT in  the  $\overline{\text{MS}} $, SI, OS and  miniMOM schemes up to order $n=9$. The variation of   $\Gamma_{\overline{\text{MS}}}/\Gamma_{\text{scheme}}$  at RG scale $\mu=\ M_H$ in the (b)FOPT and the (d)RGSPT in  the SI, OS and miniMOM schemes up to order $n=9$ . }
\label{apap_pba_scale}
\end{figure}

We summarize our predictions using the APAP formalism using the four different RG schemes in the FOPT and in the RGSPT in figure  \ref{apap_pba_scale}.  We show the the central curves of  the normalised higgs to gluons decay width at RG scale $\mu= M_H$ using four different RG schemes  in the FOPT on the top-left panel.  On the other hand,  the central curves of the ratio $\Gamma_{\overline{\text{MS}}}/\Gamma_{\text{scheme}}$ using four different RG schemes  in the FOPT are shown in the top-right panel.  The same curves in the RGSPT are shown in the bottom-left and the bottom-right panels.    We do not show similar results in the PBA formalism due to their very similarity to the APAP predictions in the $\overline{\text{MS}}$,  OS and SI schemes.

We now provide order-by-order perturbative evaluation  of $\Gamma (H \rightarrow gg)$ decay width up to the $ \rm N^5LO $ at $\mu=M_H$ in the FOPT using the PBA formalism.   The  $ \rm N^5LO $ contribution at the scale $\mu=M_H$ is highlighted inside the box.  

\begin{align}
\frac{\Gamma_{ PBA}^{\overline{\text{MS}}}(M_H)}{\Gamma_{LO}(M_H)}\,=\,&1+0.641657+0.196393+0.0176989-0.0114448 \,\, -\fbox{0.00825599},\\ \nonumber
\frac{\Gamma_{ PBA}^{SI}(M_H)}{\Gamma_{LO}(M_H)}\,=\,& 1+0.641657+0.197175+0.0178975-0.0115469-\,\, \fbox{0.00953425},\\ \nonumber
\frac{\Gamma_{ PBA}^{OS}(M_H)}{\Gamma_{LO}(M_H)}\,=\,&1+0.641657+0.196393+0.0180594 -0.0111039-\,\,\fbox{0.00914739},\\ \nonumber
\frac{\Gamma_{ PBA}^{MM}(M_H)}{\Gamma_{LO}(M_H)}\,=\,& 1+0.533291+0.0659834-0.0398094-0.0163109+\,\,\fbox{0.00114149}.
\end{align}

Similarly,  the order-by-order perturbative evaluation  of $\Gamma (H \rightarrow gg)$ decay width up to $ \rm N^5LO $ at $\mu=M_H$ in the RGSPT using the PBA formalism is,

\begin{align}
\frac{\Gamma_{ PBA}^{\overline{\text{MS}}}(M_H)}{\Gamma_{LO}(M_H)}\,=\,&1+0.695577+0.263334+0.0546671-0.00126852 \,\, -\fbox{0.0060536},\\ \nonumber
\frac{\Gamma_{ PBA}^{SI}(M_H)}{\Gamma_{LO}(M_H)}\,=\,& 1+0.686965+0.253127+0.0487537-0.00318015-\,\, \fbox{0.00739534},\\ \nonumber
\frac{\Gamma_{ PBA}^{OS}(M_H)}{\Gamma_{LO}(M_H)}\,=\,&1+0.695577+0.263334+0.055281 -0.000655924-\,\,\fbox{0.00713806},\\ \nonumber
\frac{\Gamma_{ PBA}^{MM}(M_H)}{\Gamma_{LO}(M_H)}\,=\,& 1+0.884649+0.167738-0.0114242-0.0203406-\,\,\fbox{0.00567677}.
\end{align}

\section{Determination of the Higgs to gluons decay rate}
\label{sec6}
We present our predictions of the $\Gamma (H \rightarrow gg)$ decay width in the $\overline{\text{MS}}$,  SI, OS, and miniMOM  schemes in this section.  The central value of  the $\Gamma (H \rightarrow gg)$ decay width is evaluated  at  RG scale $\mu= M_H$.       Assuming the $\Gamma_{\rm N^{n+1}LO}$ approximately to be  the exact result,  the uncertainty due to the series expansion is estimated by the difference $(\Gamma_{\rm N^{n+1}LO} - \Gamma_{\rm N^{n}LO})/\Gamma_0$ at  RG scale $\mu= M_H$.   Our predictions for the $\Gamma (H \rightarrow gg)$ decay width at the order N$^5$LO in the APAP formalism in the FOPT are,
\begin{align}
\Gamma_{\rm N^5LO}^{\overline{\text{MS}}}\,=\,& \Gamma_0\Bigl(1.837 \pm  0.047 _{\alpha_s(M_Z),1\%}\pm 0.0004_{M_t} \pm 0.0066_{M_H}  \pm  0.0009_{\rm P} \pm 0.007_{\text{s}} \Bigr), \\ \nonumber
\Gamma_{\rm N^5LO}^{\text{SI}}\,=\,& \Gamma_0\Bigl(1.837 \pm  0.046 _{\alpha_s(M_Z),1\%}\pm 0.0004_{M_t} \pm 0.0066_{M_H} \pm 0.0026_{\rm P} \pm 0.007_{\text{s}}\Bigr), \\ \nonumber
\Gamma_{\rm N^5LO}^{\text{OS}}\,=\,& \Gamma_0\Bigl(1.838  \pm 0.047 _{\alpha_s(M_Z),1\%}\pm 0.0004_{M_t}\pm 0.0066_{M_H} \pm 0.0023_{\rm P}\pm 0.007_{\text{s}}\Bigr), \\ \nonumber
\Gamma_{\rm N^5LO}^{\text{MM}}\,=\,& \Gamma_0\Bigl(1.836 \pm 0.042 _{\alpha_s(M_Z),1\%} \pm 0.0001_{M_t} \pm 0.0066_{M_H}  \pm 0.0007_{\rm P}\pm 0.0002_{\text{s}}\Bigr),
\end{align}
where $\rm P$ stands for the uncertainty due to the PBA predictions,  and  $s$ denotes the uncertainty due to the series expansion.

Similarly,   the $\Gamma (H \rightarrow gg)$ decay width using the APAP predictions at the order N$^5$LO  in the RGSPT are found to be,
\begin{align}
\Gamma_{\rm RGSN^5LO}^{\overline{\text{MS}}}\,=\,& \Gamma_0\Bigl(1.840\pm  0.047 _{\alpha_s(M_Z),1\%}  \pm  0.0005_{M_t}\pm 0.0066_{M_H} \pm 0.0002_{\mu} \pm 0.0007_{\rm P}\Bigr), \\ \nonumber
\Gamma_{\rm RGSN^5LO}^{\text{SI}}\,=\,& \Gamma_0\Bigl(1.841\pm  0.047 _{\alpha_s(M_Z),1\%}  \pm  0.0005_{M_t}\pm 0.0066_{M_H} \pm 0.0002_{\mu} \pm 0.0018_{\rm P}\Bigr), \\ \nonumber
\Gamma_{\rm RGSN^5LO}^{\text{OS}}\,=\,& \Gamma_0\Bigl(1.842\pm  0.047 _{\alpha_s(M_Z),1\%}  \pm  0.0005_{M_t} \pm 0.0066_{M_H} \pm 0.0002_{\mu} \pm 0.0019_{\rm P}\Bigr), \\ \nonumber
\Gamma_{\rm RGSN^5LO}^{\text{MM}}\,=\,& \Gamma_0\Bigl(1.847\pm  0.043 _{\alpha_s(M_Z),1\%}  \pm  0.0005_{M_t}\pm 0.0066_{M_H} \pm 0.0023_{\mu} \pm 0.0002_{\rm P}\Bigr). \\ \nonumber
\end{align}

We notice  a few important observations in our results.      The uncertainty due to the mass of the top quark in the $\overline{\text{MS}}$, SI and OS schemes turns out to be smaller,  for instance,  a change of 4 GeV in the top quark mass\cite{Alekhin:2017kpj},  causes  $0.02 \%$ change in the  $\Gamma (H \rightarrow gg)$ decay width.     In the  miniMOM scheme uncertainty due to the 4 GeV  change in the top quark mass is $0.01\%$ in the FOPT whereas it is $0.03\%$ in the RGSPT.  The dependence  of the $\Gamma (H \rightarrow gg)$ decay width on the mass of the Higgs boson   is of the order $0.36 \%$  in the $\overline{\text{MS}}$, SI, OS,  and  miniMOM scheme.   The largest uncertainty in  the $\Gamma (H \rightarrow gg)$ decay width at the $\rm N^5LO$ originates from the change in the strong coupling $\alpha_s (M_Z^2)$.   For instance,  a change of $1\%$ in the $\alpha_s (M_Z^2)$ causes an uncertainty  $(2.5-2.6) \%$ in the $\overline{\text{MS}}$,  SI and OS schemes.  For the miniMOM scheme, it is slightly less in the range $(2.3-2.4) \%$.  

As observed  in our previous discussion,  the miniMOM scheme is not showing a stable behaviour at higher orders in the APAP and PBA formalism.  Therefore,  excluding the prediction of this scheme, we  provide our final prediction of the  $\Gamma (H \rightarrow gg)$ decay width at the order N$^5$LO in the FOPT as,
\begin{align}
\Gamma_{\rm N^5LO} \,=\,& \Gamma_0\Bigl(1.8375 \pm  0.047 _{\alpha_s(M_Z),1\%}\pm 0.0004_{M_t} \pm 0.0066_{M_H} 
  \pm 0.0036_{\rm P} \pm 0.007_{\text{s}}   \pm 0.0005_{sc} \Bigr),
\end{align}
where $sc$ shows the uncertainty introduced due to scheme dependence,  and the  error due to PBA,  is obtained by adding the uncertainties due to PBA  in  the  $\overline{\text{MS}}$,  SI and OS schemes in quadrature.

In the RGSPT,  the $\Gamma (H \rightarrow gg)$ decay width at  N$^5$LO is,
\begin{align}
\Gamma_{\rm RGSN^5LO} \,=\,& \Gamma_0\Bigl(1.841 \pm  0.047 _{\alpha_s(M_Z),1\%} \pm  0.0005_{M_t}\pm 0.0066_{M_H} \pm 0.0002_{\mu} \pm 0.0027_{\rm P}  \pm 0.001_{sc} \Bigr).
\end{align}

\section{Summary}
\label{sec7}
In this work,  we have investigated an important issue of the renormalization scale and scheme dependence of  the $\Gamma (H \rightarrow gg)$ decay width in the FOPT and in the RGSPT at the order  $\rm N^4LO$,  and beyond it.   The RGSPT exploits the method of summation of  all RG-accessible logarithms,  which was first proposed  in reference \cite{Maxwell}.  In the RGSPT,  the RG equation is utilized to derive the summation of the leading and subsequent finite subleading logarithms  to all orders in the perturbation theory.  This results in a closed-form summation of the RG accessible  leading and subsequent finite subleading logarithms.   It is found that the dependence of the perturbative expansions on the RG scale $\mu$ is considerably reduced in the  RGSPT expansions.

This work is motivated by a recent advancement in the computation of the $H \rightarrow gg$ decay rate in the limit of a heavy top quark and any number of massless light flavours at N$^4$LO in reference~\cite{Herzog:2017dtz}.    We investigate the $\Gamma (H \rightarrow gg)$ decay width in four different renormalization schemes,  namely,  $\overline{\text{MS}}$,  SI,  OS, and miniMOM.  We first discuss our predictions in the FOPT,  and then compare these predictions to those  obtained in the RGSPT.  In the case of the FOPT expansions,  the $\Gamma (H \rightarrow gg)$ decay width is highly sensitive to  the RG scale up to the order $\rm N^2LO$,  and stabilizes at the order $\rm N^4LO$.    We observe that the summation of leading logarithms in the RGSPT expansions exhibits good stability and reduced sensitivity to RG scale $\mu$.  This is starkly obvious up to the order $\rm N^2LO$.  The FOPT expansions begin to catch up with the behaviour of the RGSPT expansions from $\rm N^3LO$ onward.   Moreover,  the RGSPT expansions  show a stable behaviour in different RG schemes as well.   The largest uncertainty in our predictions for the $\Gamma (H \rightarrow gg)$ decay width arises due to the $1\%$ change in the strong coupling $\alpha_s (M_Z^2)$, and is in the range $(2.5-2.6) \%$ in the $\overline{\text{MS}}$,  SI and OS schemes.  The corresponding range in the miniMOM scheme is $(2.3-2.4) \%$,  which is slightly less than those obtained in the  $\overline{\text{MS}}$,  SI and OS schemes.   

We have also estimated the higher-order effects on the $\Gamma (H \rightarrow gg)$ decay width  using the APAP formalism.  The higher-order behaviour of the perturbative expansions is alternatively determined by the PBA formalism,  and found to be reasonably  in agreement with that of the APAP formalism  for the   $\overline{\text{MS}}$,  SI,  and OS schemes.     The  $\Gamma (H \rightarrow gg)$ decay width is showing  stability  at higher-orders  in the APAP as well as in the PBA frameworks,  and it becomes less dependent on the higher-order corrections in all  schemes.   The RGSPT expansions continue to show greater stability against the RG scale  at higher-orders in the APAP as well as  in the PBA frameworks.

Finally,  we provide our estimate of the $H \rightarrow gg$  decay rate at N$^5$LO  in the framework of the FOPT as well as the RGSPT.  We have added the difference between the APAP and the PBA  predictions of the  $H \rightarrow gg$  decay rate at N$^5$LO as an error to the final predictions of the $H \rightarrow gg$  decay rate.  This uncertainty is approximately $0.19\%$  in the FOPT,  and  $0.15\%$  in the RGSPT.   The uncertainty due to the truncation of series is approximately  $0.6\%$ at N$^4$LO,  and reduces to $0.4\%$ at N$^5$LO in the FOPT.    Thus,  by adding N$^5$LO correction to the  $H \rightarrow gg$  decay rate, the truncation error is reduced by $33\%$ at N$^5$LO.   The uncertainty due to  the truncation  is  much smaller than the error introduced by the $1\%$ uncertainty in the strong coupling constant.  We notice that the uncertainty of the order $1\%$ in the  $\Gamma (H \rightarrow gg)$ decay width  may not be in the reach of  the LHC.  However,  this precision may be accessible  to  a future $e^+ e^-$ linear collider\cite{Fujii:2015jha}.

We emphasize that the uncertainty due to the missing higher-order effects of QCD corrections beyond N$^3$LO is an important issue in the Higgs physics for the upcoming high-luminosity phase of the LHC \cite{Cepeda:2019klc}.  This uncertainty reveals itself in the form of scale-dependence.  The uncertainty due to scale is remarkably reduced in the framework of the RGSPT,  and is approximately  $0.01\%$.    Our final predictions for the $H \rightarrow gg$  decay rate at N$^5$LO in the FOPT and in the RGSPT includes the uncertainty entering due to the scheme dependence,   which  is  approximately $0.03\%$ in the FOPT,  and   $0.06\%$  in the RGSPT.     This prediction is obtained by excluding the  miniMOM scheme,  which is  not working well within  the APAP and PBA formalisms,  and requires further future investigation.

There are other sources of uncertainties to  the $\Gamma (H \rightarrow gg)$ decay width.  For instance,  the electroweak corrections cause the enhancement of  the $\Gamma (H \rightarrow gg)$ decay width by about $5\%$ \cite{Degrassi:2004mx,Aglietti:2006yd,Actis:2008ug,Actis:2008ts}.  The missing electroweak corrections beyond $\rm NLO$ introduce the residual theoretical uncertainties  of the order $1\%$ \cite{Denner:2011mq}.  Additionally,  corrections due to  a finite bottom-quark mass induce a $12\%$ effect at leading order, and $6\%$ effect at $\rm NLO$ to the  effective Higgs coupling to gluons \cite{Mueller:2015lrx,Lindert:2017pky}.  A comprehensive analysis of these effects in the framework of the RGSPT is beyond the scope of this paper,  and will be presented in future work.

\section*{Acknowledgments}

We are extremely grateful to the referee for the highly constructive feedback on this work.   We are also grateful to Prof.  Irinel Caprini,  Prof. B. Ananthanarayan and M.  S.  A.  Alam Khan for very important comments and suggestions on the manuscript.  We are also very thankful to Prof.  M. Spira for very useful comments and suggestions  on the first arXiv version of the manuscript.  This work is supported by the  Council of Science and Technology,  Govt. of Uttar Pradesh,  India through the  project ``   A new paradigm for flavour problem "  no.   CST/D-1301,  and Science and Engineering Research Board,  Department of Science and Technology, Government of India through the project `` Higgs Physics within and beyond the Standard Model" no. CRG/2022/003237. 
\section*{Data availability statements}
The used data is explicitly quoted in the manuscript itself,  and there is no need to deposit it separately.

\begin{appendix}

\end{appendix}

\end{document}